\documentclass[11pt]{article}

\usepackage[a4paper,margin=1in]{geometry}
\usepackage{times}
\usepackage{amsmath,amssymb}
\usepackage{booktabs}
\usepackage{longtable}
\usepackage{microtype}
\usepackage{enumitem}
\usepackage{cite}
\usepackage{xcolor}
\usepackage{graphicx}
\usepackage[colorlinks=true,allcolors=blue]{hyperref}

\title{Peripheral Physiological Decoding of Arousal in a Boundary-Avoidance Task: Signal Validity and Individualized Control}
\author{Anirudh Natarajan$^{1}$ and Paul Sajda$^{1,2,3,4}$\\[1ex]
\small $^1$Department of Biomedical Engineering, Columbia University, New York, NY 10027, USA\\
\small $^2$Department of Electrical Engineering, Columbia University, New York, NY 10027, USA\\
\small $^3$Department of Radiology, Columbia University Irving Medical Center, New York, NY 10032, USA\\
\small $^4$The Data Science Institute, Columbia University, New York, NY 10027, USA\\[1ex]
\small E-mail: \href{mailto:amn2225@columbia.edu}{amn2225@columbia.edu}}
\date{}

\begin{document}
\maketitle

\begin{abstract}
\textbf{Objective.} Scoring well on a task-demand label does not show that a decoder measures arousal. We ask whether peripheral physiology decodes demand better than EEG, whether the index behaves like graded arousal rather than a clock, and what a controller could do with it.

\textbf{Approach.} We re-analysed a public boundary-avoidance flight task in which sixteen participants received EEG neurofeedback, half-sham feedback or silence. A decoder trained on peripheral signals (cardiac, respiratory, electrodermal, pupillometric) from an open-loop calibration block was compared with a retrained EEG decoder, filter-bank common spatial patterns, and motor-only and shuffled-label controls. All analyses are offline; the decoder was never in the loop.

\textbf{Main results.} The peripheral decoder reached $93.0\%$ within-subject AUC against $85.2\%$ for EEG. Ring size shrinks on a fixed schedule, so accuracy alone cannot separate arousal from trial position. Three tests did: $48.6\%$ of hard-course trials ended below the scale midpoint; the index separated the courses at matched elapsed time ($p = 1.4\times10^{-31}$), where EEG did not; and it tracked the autonomic channels. Only the peripheral index recovered an inverted-U with performance, and it held in a duration-safe hazard analysis ($p = 0.043$). Arousal rose $14.1$ points in the $10$\,s before failure; sustained between-condition differences above $7.5$ points are excluded. Baseline measures from pre-loop calibration predicted the shape of a subject's arousal-performance curve. Personalised control bands gave the best held-out discrimination, though at $N = 16$ indistinguishable from a retuned universal band.

\textbf{Significance.} Arousal is a state of the whole body, and here the body reads it better than the scalp. The results separate three often-conflated requirements: accuracy on a demand label, validity as an arousal signal, and usefulness for control. The signal a controller should act on is graded distance above the band rather than a threshold crossing.
\end{abstract}


\section{Introduction}

The Yerkes-Dodson law \cite{yerkes1908, diamond2007} posits an inverted-U relationship between arousal and cognitive performance: performance is maximized within an optimal arousal band and degrades on either side. Maintaining an operator within that band is a compelling target for closed-loop brain-computer interfaces (BCIs), with direct relevance to high-stakes domains where arousal escalation leads to catastrophic failure, such as pilot-induced oscillations in aviation \cite{gray2005, saproo2016}.

Faller et al \cite{faller2019} demonstrated a proof of concept using a virtual-reality boundary-avoidance task (BAT). An EEG decoder based on filter-bank common spatial patterns (FBCSP) \cite{ang2012} and linear discriminant analysis decoded task-dependent arousal with an AUC of $79.8\%$ and drove loudness-modulated auditory neurofeedback. Subjects receiving veridical feedback flew approximately 7\,s longer on the hard course than in silence, with corroborating increases in HRV and reductions in pupil diameter suggesting genuine down-regulation of arousal. That work remains a primary demonstration that EEG-based neurofeedback can improve performance in a demanding real-time task.

We re-analyse the same dataset, under two limits that apply throughout. Every analysis here is offline. The peripheral decoder we develop was never in the feedback loop, so the performance benefit Faller et al report cannot be attributed to it, and the closed-loop architecture in section~\ref{sec:framework} is a specification, not a working system. And the decoder is trained on a task-demand label derived from ring size, not on an independent measure of arousal. Recovering the expected arousal-performance relationship from such a decoder is evidence within one paradigm, not proof that the signal is arousal.

Within that scope we address three gaps. First, decoding accuracy in the original study was modest, and the continuous output of the FBCSP decoder was never checked against the arousal-performance relationship it was meant to track. High classification accuracy does not guarantee that a decoded signal represents graded arousal: a decoder trained to separate large from small rings can score well on any feature correlated with task difficulty. Second, although feedback improved hard-course flight time, the mechanism by which it acts on the arousal signal was not characterised. Does it shift the overall arousal level, or does it act locally at performance-critical moments? Third, feedback was delivered reactively against a uniform threshold, ignoring documented individual differences in arousal sensitivity.

A broader question motivates all three. Why should EEG be the default modality for arousal decoding? Arousal is usually treated as a cognitive state, and cognitive states are measured at the scalp. We start elsewhere. Arousal is a state of the whole body. The locus coeruleus-norepinephrine (LC-NE) system \cite{astonjones2005} drives cortical excitability and peripheral autonomic output together, so a rising heart rate, a dilating pupil and a skin conductance response are not downstream echoes of something happening in cortex. They are that state showing itself \cite{bradley2008}. The body also shows it early and at high signal-to-noise, while at the scalp the same shift has to be recovered from a far noisier mixture. Arousal matters for cognition, and it matters for a BCI. But a decoder that reads only the head reads one surface of a state the rest of the body is already reporting. Peripheral signals are heavily used in affective computing \cite{picard2001, schmidt2018} and almost absent from BCIs, despite being closer to the target variable and easier to deploy. EEG may well retain value elsewhere; the question here is which modality carries the task-relevant information in this paradigm.

A closed-loop arousal BCI needs three things from its decoder. It must be
accurate. Its output must behave like the state it claims to measure. And that
output must expose something a controller can act on. These are usually treated
as one requirement; clearing the first does not clear the second, and the three
Results sections take them in turn.

A multimodal decoder using peripheral signals beats both an EEG-only variant and
the original FBCSP benchmark within subject. Across subjects the comparison is
too weak to confirm or deny the same ordering. Accuracy alone settles little
here in any case, because the label is almost a deterministic function of
position within the trial.

Four results address validity. The index is not a trial clock. It separates the
two courses at matched trial phase, where a matched-scaling EEG index does not.
It tracks raw autonomic signals in the expected directions. And it recovers the
inverted-U with performance, where the EEG and motor indices do not. We then ask
what feedback does to it, using a failure-locked analysis and an explicit bound
on what the trajectory null can exclude.

On control, excess arousal above the band predicts imminent failure, though at
$N = 16$ the differences between subjects in how strongly it does cannot be
estimated. Acting on it is harder than it looks. Scaling the width of the
control band by a per-subject measure cannot work, for a structural reason we
set out, and no per-subject scheme we tested beat a universal band retuned to
one multiplier for everybody. Two parameters escape that argument: the centre of
the band, and the graded distance above a band set by the subject's own recent
variability. What matters is the graded form rather than the reference.

% ═══════════════════════════════════════════════════════════════════════════════
\section{Data and experimental setup}

We analysed data we previously collected and made publicly available in Faller et al \cite{faller2019} (IEEE DataPort: \url{https://doi.org/10.21227/rn3e-bp31}). Full experimental details are given in that paper; we describe here only what is necessary to interpret the present analyses.

\subsection{Participants and task}

$N = 16$ healthy adults performed a virtual-reality boundary-avoidance task in which they navigated a simulated aircraft through sequences of red rectangular boundary rings arranged along a sinusoidal trajectory. Ring sizes decreased every 30\,s (large, medium, small), increasing task difficulty across each trial of at most 90\,s. Performance was measured as time in flight before missing a ring. Two courses were used. The easy course used ring diameters of 108, 72 and 36 units for large, medium and small rings; the hard course used 60, 30 and 14.4 units. Most analyses focus on the hard course, whose difficulty range drives arousal into the region where Yerkes-Dodson dynamics are most pronounced. Because the two courses differ in ring geometry at the same elapsed time, they also make the matched-phase test of section~\ref{sec:validity} possible.

The closed-loop experiment comprised alternating easy and hard flights. Each flight was randomly assigned to one of three feedback conditions: BCI (an EEG-decoded arousal index modulating the loudness of auditory heartbeat feedback); half-sham (a 50/50 blend of the true decoder output and an autoregressive surrogate); or silence (no feedback). Conditions rotated so that each appeared twice per six consecutive flights. Half-sham delivers half of the veridical signal, so it is a partial rather than a pure control; section~\ref{sec:deviation} reports what including it costs.

\subsection{Calibration block}

Decoder training used a separate 10-minute open-loop calibration session recorded before the closed-loop experiment: 136 trials, all on the easy course, in silence. The decoder therefore never sees hard-course ring geometry during training, and transfer to the analysed trials is across session, course, ring geometry and feedback condition.

\subsection{Physiological recordings}

All signals were recorded at 2048\,Hz (BioSemi ActiveTwo) and downsampled to 256\,Hz prior to analysis. Recordings comprised 64-channel EEG, ECG from two thoracic electrodes, EDA from two left-hand electrodes, respiration from a chest belt, and pupillometry from the VR headset eye tracker (60\,Hz), synchronized via Lab Streaming Layer \cite{kothe2013}. From these raw signals we derived seven peripheral channels: heart rate (HR), beat-to-beat HRV (pNN35) \cite{taskforce1996}, respiratory amplitude (RESP), phasic EDA, tonic EDA, and left and right pupil diameters (PUP-L, PUP-R). Joystick and head-motion channels were retained for the behavioural control decoder. No preprocessing beyond downsampling was applied to peripheral channels prior to decoder training.

\subsection{Epoch extraction and labels}\label{sec:labels}

For each ring crossing event, a 2\,s (512-sample) epoch of physiological signal immediately preceding the crossing was extracted. Following Faller et al \cite{faller2019}, labels were assigned by ring size: large rings (low task demand) were class 0; medium and small rings (high task demand) were class 1. Training data comprised epochs from the calibration session; held-out test data comprised all epochs from the closed-loop phase.

One property of the design limits what accuracy on this label can show, and it is worth stating before any accuracy is reported. Rings shrink on a fixed schedule, so the small rings always fall late in a trial. The label is therefore close to a deterministic function of position within the trial, and ring position alone predicts it at $99.8\%$ AUC. This rules out the two controls one would normally reach for: a phase-stratified evaluation is undefined, because almost no ring position contains both classes, and shifting the labels in time changes too few of them to test anything (supplementary table~S4). Section~\ref{sec:validity} therefore settles validity with comparisons that vary ring size independently of elapsed time.

% ═══════════════════════════════════════════════════════════════════════════════
\section{Methods}

\subsection{Multimodal peripheral decoder}

\subsubsection{Architecture.}
The seven peripheral channels were grouped into four functional modalities: cardiac (HR, HRV-pNN35), respiratory (RESP), electrodermal (EDA-phasic, EDA-tonic), and pupillometric (PUP-L, PUP-R). Each modality was independently processed by a \textit{SimpleTemporalEncoder} consisting of a two-layer 1D convolutional stack (kernel size 3, same padding, batch normalization, ReLU) followed by global average pooling into a 64-dimensional embedding. The four embeddings were concatenated into a 256-dimensional vector and classified by a two-layer MLP (hidden dimension 128, ReLU, dropout $p = 0.1$, output dimension 2). All weights used Kaiming normal initialization. The architecture is illustrated in figure~\ref{fig:figure_1}A.

\subsubsection{Training.}
Per-channel z-score normalization was computed from training data only and applied to held-out test data. Data augmentation applied two strategies independently with $p = 0.5$: temporal masking (approximately 15\% of the epoch zeroed at a random contiguous position) and additive Gaussian noise ($\sigma = 0.1$). Cross-entropy loss with label smoothing ($\varepsilon = 0.05$) was minimized with Adam ($\text{lr} = 10^{-3}$), using a step learning rate scheduler ($\gamma = 0.1$, every 10 epochs) for up to 25 epochs with early stopping (patience 10, activated after epoch 15). The autonomic decoder was trained 10 times per subject with independent random seeds; the EEG-only and combined decoders were trained 5 times. Softmax probabilities were averaged across repetitions.

\subsubsection{Continuous arousal estimation.}\label{sec:scaling}
For trial-level analysis, the trained model was applied to a sliding window advancing every 16 samples (62.5\,ms), using the most recent 512-sample epoch. Raw probabilities were normalized to a 0--100 index using the 5th and 95th percentiles of that subject's training distribution, then clipped to that range. The result was filtered with an asymmetric first-order infinite impulse response filter ($\tau_\text{rise} = 0.75$\,s, $\tau_\text{fall} = 1.5$\,s), which reflects the faster escalation than recovery of autonomic arousal. The time constants match established sympathetic activation latency (under 1\,s for HR acceleration) against parasympathetic recovery (1--3\,s) \cite{berntson1997}. The filter is invertible, so a stored trace can be unsmoothed and re-smoothed with any other constants without retraining a decoder. Nothing downstream depends on the choice. Under symmetric filters at $\tau = 0.75$, $1.0$ and $1.5$\,s, and with no smoothing at all, the deviation-performance correlation stays between $-0.273$ and $-0.314$ against $-0.308$ for the filter used here, and the trial-level quadratic between $-2.30$ and $-2.50$ against $-2.33$. Every value in that comparison is recomputed from the resmoothed trace, which is why the reference differs slightly from the $-2.56$ that section~\ref{sec:yerkes} reports from the stored trial means. Supplementary table~S5 gives the full comparison.

The normalization has three properties that matter downstream. It is subject-specific, so index values compare within a subject but not across them. It is clipped, so values saturate at the ends of the range. And it differs from the scaling applied to the comparison signals: the EEG feedback channel \texttt{FB-HB-0-1-nrm} and the raw physiological traces are min-max scaled within each trial, which removes between-trial differences in level by construction. Trial-level comparisons between the two indices therefore use matched per-subject scaling as primary, with the original pipeline's per-trial scaling reported alongside.

\subsubsection{Comparison conditions and controls.}
The peripheral decoder was compared against three alternatives: a retrained EEG-only variant \cite{lawhern2018}, the original FBCSP benchmark, and a combined autonomic and EEG variant. The EEG-only variant used the identical architecture applied to 64-channel EEG decomposed into four standard frequency bands, giving 256 input channels (theta 4--8\,Hz, alpha 8--13\,Hz, beta 13--30\,Hz and gamma 30--55\,Hz, via Butterworth bandpass filters). Two controls used the same architecture and recipe: a motor decoder taking joystick and head-motion channels only, and a shuffled-label control in which labels were permuted within subject. Feature attribution used Integrated Gradients \cite{sundararajan2017} with 50 integration steps, averaged across held-out validation epochs.

Cross-subject generalization was assessed by leave-one-subject-out validation: 16 folds, each training on the calibration epochs of 15 subjects and testing on the sixteenth, 3 repetitions per fold, with channel normalization fitted on the training subjects only.

\subsection{Validity of the decoded index}\label{sec:validity_methods}

Three tests address whether the index carries trial-specific arousal information rather than tracking position within a trial.

\subsubsection{Endpoint distribution.} A signal that is a function of elapsed time must end each trial near the top of its own ramp. We therefore examined the distribution of decoded arousal in the final 3\,s of each hard-course trial, and its relationship to trial duration.

\subsubsection{Matched-phase course discrimination.} Within a trial, ring size and elapsed time are confounded. Between courses they are not: the hard course uses smaller rings than the easy course at every elapsed time. We fitted mixed models of decoded arousal at each ring crossing on course (hard versus easy) with elapsed time as a covariate (linear, and linear plus quadratic), and a variant restricted to large rings only, which holds the nominal label fixed. A signal driven purely by elapsed time predicts a course coefficient of zero. The same models were fitted to the EEG index at both scalings and to the motor index.

\subsubsection{Convergent physiological validation.} Decoded arousal was correlated with each raw physiological channel within each trial, and the per-trial coefficients averaged; significance was assessed with a one-sample test across subject means, with a rank check alongside. Seven ways of computing the same correlation are compared in supplementary table~S3, because they differ substantially and the choice needs stating.

\subsection{Arousal and performance}

To test the inverted-U we fitted linear mixed-effects models of the form
\begin{equation}
  \text{Perf} \sim \text{Ar} + \text{Ar}^2 + \text{Diff} + \text{Cond} + (1 \mid \text{Subj}),
  \label{eq:lme}
\end{equation}
where $\text{Ar}$ is the average decoded arousal for the trial, Diff is ring difficulty and Cond is feedback condition. A negative coefficient on Ar$^2$ indicates the inverted-U. Fits were compared across decoders using the Akaike information criterion.

Trial-mean arousal is averaged over a window whose length is the outcome, so predictor and outcome are not independent and the model-based $p$-value is anticonservative. We therefore report a permutation $p$-value obtained by shuffling arousal traces between trials within duration strata (400 shuffles), which preserves the duration coupling exactly while destroying the pairing between a trial and its own arousal.

Two further analyses avoid the length coupling altogether. A landmark analysis is the standard survival-analysis device for keeping predictor and outcome on separate stretches of a trial: fix a landmark time $L = 30$\,s, keep only the trials still flying at $L$, measure arousal in the 5\,s window ending at $L$, and predict the flight time remaining after $L$. The name refers to the landmark time, not to the size of the result. A discrete-time hazard model instead takes the 1\,s bin as the unit of analysis and asks whether current arousal predicts crashing within the next 5\,s, controlling for elapsed time with 5\,s bin fixed effects. The outcome is binary and its windows overlap across consecutive bins, so every hazard result is refitted as a binomial GEE with standard errors clustered on trial, and again on non-overlapping 5\,s blocks. Supplementary section~S9 gives both.

\subsection{Effect of feedback condition}

Mean arousal traces were compared across BCI, half-sham and silence trials with a cluster-based permutation test (parameters in section~\ref{sec:stats}). Because the headline result is a null, we also estimated what size of sustained condition difference the design could have detected, by bootstrapping the observed subject-level differences with an injected constant offset and rerunning the same test.

Separately, arousal in the final 10\,s before a crash was compared with the preceding 10\,s window, overall and by condition, and the hazard model was extended with an interaction between above-band arousal and feedback condition.

\subsection{Individual differences}\label{sec:individual_methods}

Every baseline measure comes from the calibration block alone, so that all of them would be available before a loop closes. Seventeen measures were screened for moderation of the arousal-performance curve, by adding $\text{Measure}_z \times \text{Ar}^k$, $k \in \{1,2\}$, to equation~(\ref{eq:lme}). Twelve were physiological: HRV-pNN35 and its within-session variability, respiration, heart rate, phasic and tonic EDA, pupil diameter, four EEG band powers (theta, alpha, beta and gamma) and the theta/beta ratio. Five were statistics of the decoder's own out-of-fold output during calibration: level, standard deviation, interquartile range, volatility and time spent high. The seventeen are listed with their coefficients in supplementary table~S2. The screen was Holm-corrected across its 17 members and repeated leave-one-subject-out to check that no single subject carried a result.

HRV-pNN35 was singled out ahead of the screen on theoretical grounds: the Thayer et al \cite{thayer2012} meta-analysis ties resting HRV to prefrontal regulatory capacity, which predicts a flatter arousal-performance curve. We therefore treat it as a single a priori hypothesis rather than a member of the 17-measure screen, report it uncorrected, and label it as such wherever it appears (section~\ref{sec:stats}). Because HRV-pNN35 has a single definition, it cannot be moved by the choice of estimator that retired the composite score; treated instead as one member of the screen it would not survive correction (Holm $p = 0.26$), and we report that value too. EEG band power in the screen is computed as mean square power throughout.

\subsection{Optimal arousal trajectory and control}

\subsubsection{Trajectory derivation.}
Using hard-course trials under silence and half-sham conditions (113 trials, that is every hard-course control trial, not the subset with a 20\,s pre-crash window used in section~\ref{sec:feedback}), mean performance was computed for each combination of 1-second time bin and 5-unit arousal bin. Bins were smoothed with a Gaussian kernel. The optimal arousal $\hat{a}(t)$ at each time point was taken as the performance-weighted mean arousal over sufficiently populated bins, and the half-width $\hat{\sigma}(t)$ as the weighted standard deviation around it. Both series were fitted with a cubic polynomial weighted by bin occupancy. We use \textit{control band} to mean the interval $[\hat{a}(t) - m\hat{\sigma}(t),\; \hat{a}(t) + m\hat{\sigma}(t)]$ for a width multiplier $m$; the \textit{universal} band uses $m = 1$ for every subject, and a \textit{group-specific} band uses a different $m$ for each sensitivity group.

\subsubsection{Deviation metrics.}
For each trial we computed the percentage of time above the upper bound, the percentage within the band, the mean above-band excursion magnitude, and the number of band crossings per minute. The first, second and fourth are monotone transforms of one another, so only two independent quantities are reported (section~\ref{sec:deviation}). We also measured how long an above-band excursion lasts before arousal returns, which sets the latency budget for a reactive intervention; excursions shorter than 2\,s were discarded as transients.

\subsubsection{Held-out evaluation of personalized bands.}
Band width multipliers were optimized by grid search over $m \in \{0.50, 0.75, \ldots, 3.00\}$. Selecting and evaluating on the same trials would overstate the result, so the primary evaluation is leave-one-subject-out. By \textit{group split} we mean dividing subjects at the median of a calibration measure into an arousal-sensitive and an arousal-tolerant half, and giving each half its own multiplier. Per fold, the trajectory, the median and both multipliers were refitted on the 15 training subjects, and the held-out subject was assigned to a group by its own calibration value against that training median, so nothing from the held-out subject's trials enters the fit. A second evaluation applied parameters fitted on the silence and half-sham trials, unchanged, to the hard-course full-BCI trials, which entered no part of the fit.

\subsubsection{Reliability of the quantities being personalized.}
Personalizing a parameter only makes sense if the parameter is stable within a person. We estimated each subject's optimal arousal and best band width independently in random halves of their own trials (200 splits) and correlated the halves across subjects. A subject's optimum tracks their mean arousal closely, so we recomputed the same reliability after residualizing the optimum on that mean.

\subsubsection{Reactivity to excess arousal, and a band derived from the fitted curve.}
Band width is a constant within a subject, and a constant cannot reorder that
subject's own moments, so we also asked whether subjects differ in something a
controller could act on second by second. We define reactivity as the slope
relating graded excess above the band to the probability of crashing within
5\,s, and test it by adding a random slope on that excess to the hazard model.
To ask whether a baseline measure predicts reactivity, we add an interaction
between the measure and the excess to the same model.

We then derived each subject's band from their own fitted arousal-performance
curve instead of by grid search. If performance falls as $c(a - \hat{a})^2$, a
fixed tolerated loss implies a half-width proportional to $1/\sqrt{|c|}$, so a
steeper curve gives a tighter band. Curvature and optimum come from the
interaction model of section~\ref{sec:individual_methods}, evaluated at the
held-out subject's calibration measure. This yields two parameters that can be
varied separately, the width and the centre. Everything is fitted
leave-one-subject-out, including the trajectory and the interaction, and the
comparison includes a universal band at the best single multiplier, because a
personalized scheme has to beat a retuned global one, not just the default.

\subsubsection{Within-trial control parameters.}
The hazard model asks the question a controller faces: given the current second, does this signal predict a crash within the next 5\,s? Eight candidate signals were compared in that framework. Four are threshold rules: the fixed universal band, a retuned universal band, a per-subject band scaled by calibration variability, and an adaptive band whose width comes from the subject's own arousal variability in a trailing 10\,s window. The remaining four are a sustained variant requiring 3 consecutive seconds above band, the graded excess above the fixed band, the graded excess above the adaptive band, and the rate of change of arousal. Discrimination is reported as AUC computed within 5\,s time bins, together with the mixed-model coefficient and a likelihood-ratio test against a time-only null, and each candidate was also tested for added information over the fixed rule.

\subsection{Statistical analysis}\label{sec:stats}

Each participant contributed many trials, so trial-level Pearson and Spearman correlations are reported as descriptive quantities and every inferential statement uses a linear mixed-effects model with a random intercept per subject. Coefficients are reported with standard errors, 95\% confidence intervals, $p$-values and trial and subject counts. Mixed models were fitted by maximum likelihood using statsmodels in Python 3, which AIC comparison across different fixed-effect structures requires. Models with a random slope on arousal are reported where they converge; a random slope on arousal$^2$ did not converge and is not reported.

The cluster-based permutation test is specified as follows. The permutation unit is the subject, with condition labels permuted within subject, because trials are nested in subjects. The observation is one mean trajectory per subject per condition, so each subject contributes equally regardless of how many trials they completed. The pointwise statistic is a repeated-measures $F$ across the three conditions, and a paired $t$ for pairwise contrasts. The cluster-forming threshold is the $p < 0.05$ critical value of that statistic ($F = 3.32$, $t = 2.13$ on the hard course). The cluster statistic is the sum of the pointwise statistic over contiguous supra-threshold bins, and the null distribution is the maximum cluster statistic per permutation, which controls the family-wise error rate across time. We used 10{,}000 permutations over 30 one-second bins, dropping the first two: the first second of every trace is exactly zero by construction and the filter ramps through the second, so those bins have no between-subject variance. Every cluster and every threshold is listed in supplementary section~S8.

Multiple comparisons follow one policy, applied to every $p$-value that supports a claim in the Results: 75 tests across 15 families. A family is a set of tests addressing one question with interchangeable outcomes, and Holm correction is applied within family. An a priori primary test forms a family of one, so HRV-pNN35, singled out ahead of the screen on theoretical grounds (section~\ref{sec:individual_methods}), is reported uncorrected and marked as such, whereas respiration emerged from the 17-measure screen and carries its corrected value. HRV also appears inside that screen, where it does not survive; we give both values rather than the more convenient one. Descriptive quantities carry no $p$-value and enter no family. Cluster-level $p$-values already control error across time within a test, so the family there is the set of tests, not the set of bins. Where a result sits close to the threshold we give the raw and the adjusted value. Diagnostic and sensitivity tests qualify a claim rather than support one, so they enter no family: these are the alternative specifications of the matched-phase model, the refits on narrower control-trial subsets and the in-sample oracle, each marked where it appears. Nothing here was preregistered. This is a re-analysis of an existing dataset, and a priori means specified on theoretical grounds before the relevant outcomes were examined. The full registry is supplementary table~S1: 75 tests, 45 of them significant uncorrected, and 12 of those 45 do not survive their family's correction.

\begin{figure}[!htbp]
  \centering
  \includegraphics[width=\linewidth]{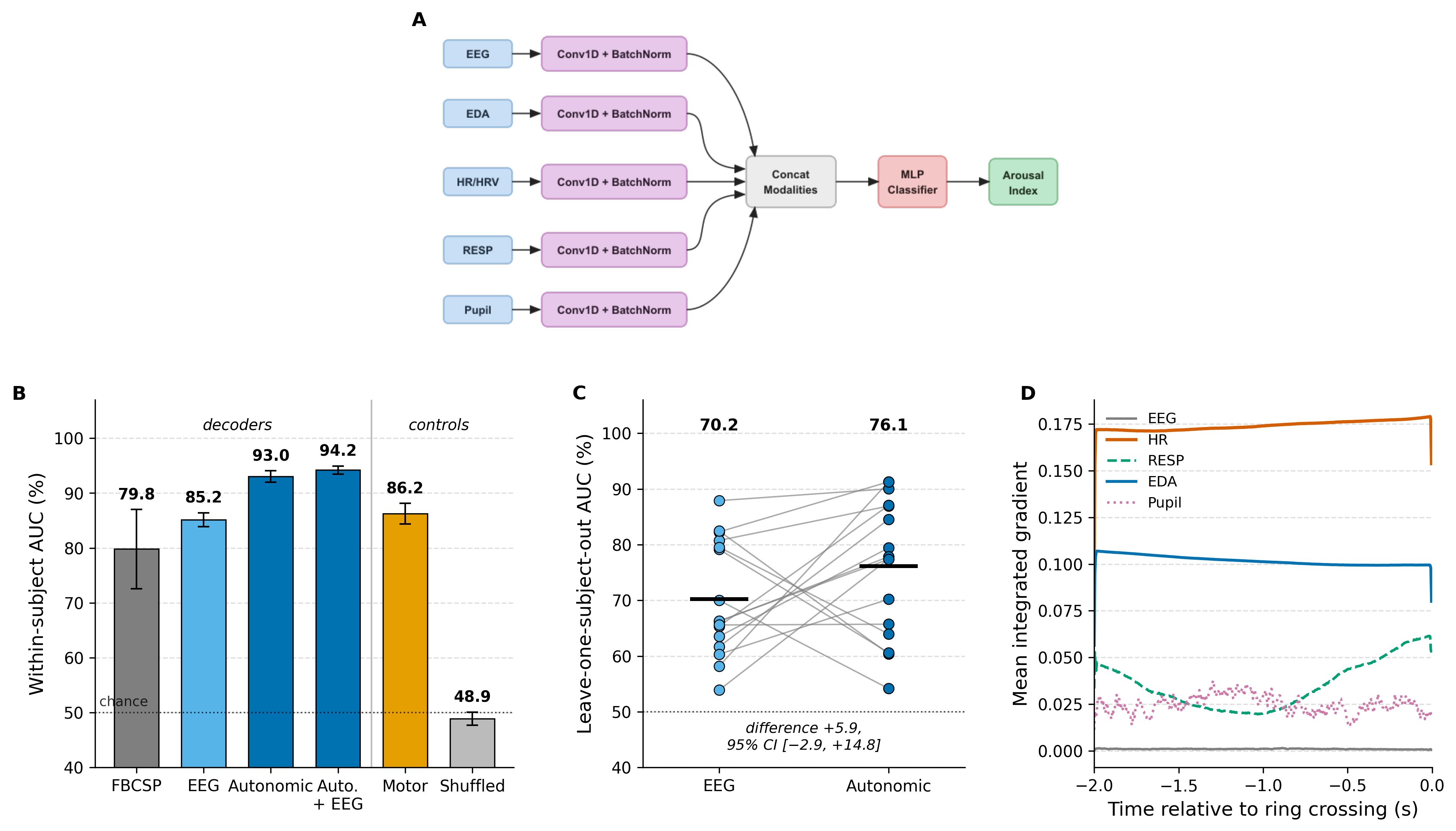}
  \caption{Decoder accuracy, cross-subject generalization and attribution. (A) Network architecture: each input modality is independently embedded by a Conv1D and BatchNorm temporal encoder, embeddings are concatenated, and a two-layer MLP produces the arousal index. (B) Within-subject AUC for the original FBCSP benchmark, the retrained EEG decoder, the autonomic decoder and the combined decoder, alongside two controls run through the identical pipeline: a motor decoder using joystick and head motion, and shuffled labels. Error bars are SEM across the 16 subjects. (C) Leave-one-subject-out AUC for the EEG and autonomic feature sets, one point per held-out subject, with subjects joined across feature sets and group means shown as horizontal bars. (D) Integrated Gradients attribution by modality across the 2-second pre-event epoch in the combined model. Cardiac signals dominate, EDA contributes substantially, respiration rises late in the epoch, and EEG attribution is near zero throughout. Throughout, ``decoded arousal'' denotes the output of a decoder trained on a task-demand label; its interpretation as arousal rests on the validity tests of section~\ref{sec:validity} (see section~\ref{sec:validity_discussion}).}
  \label{fig:figure_1}
\end{figure}

% ═══════════════════════════════════════════════════════════════════════════════
\section{Results}

\subsection{Decoding accuracy and its limits}

The autonomic decoder achieved within-subject AUC of $93.0\% \pm 1.1\%$ and the combined autonomic and EEG variant $94.2\% \pm 0.71\%$, both above the retrained EEG-only decoder ($85.2\%$) and the original FBCSP benchmark ($79.8\% \pm 7.2\%$; figure~\ref{fig:figure_1}B). Autonomic exceeded EEG by 7.7 points ($p = 8.7\times10^{-6}$, paired $t$-test) and combined exceeded EEG by 9.0 points ($p = 4.2\times10^{-7}$). Adding EEG to the autonomic decoder gained $+1.2$ points ($p = 0.016$), which is small and only nominally significant: EEG carries little that the peripheral channels do not already have. All 16 subjects exceeded $80\%$ AUC under the autonomic decoder and 12 of 16 exceeded $90\%$. Subsequent analyses use the autonomic decoder, with the combined variant retained for sensitivity checks.

Feature attribution says the same thing. Cardiac signals dominated, tonic EDA contributed substantially, respiration and pupillometry made modest but stable contributions with a late-epoch rise in respiratory attribution in the final 500\,ms before ring crossing \cite{joshi2016}, and EEG attribution was near zero throughout the epoch even in the combined model (figure~\ref{fig:figure_1}D).

\subsubsection{The modality advantage does not replicate across subjects.}
Under leave-one-subject-out validation the autonomic decoder reached $76.1\% \pm 3.05\%$ (range $54.2$ to $91.3$) and the EEG decoder $70.2\% \pm 2.57\%$ (range $53.9$ to $87.9$); both are above chance ($p = 3.7\times10^{-7}$ and $p = 1.1\times10^{-6}$). The autonomic figure is 16.8 points below the within-subject value ($t = 6.35$, $p = 1.3\times10^{-5}$), and three held-out subjects fall below 61. The peripheral advantage over EEG was $+5.9$ points with a 95\% confidence interval of $[-2.9, +14.8]$ ($p = 0.173$; figure~\ref{fig:figure_1}C). Twelve of 16 subjects favoured the peripheral decoder, and the effect points the same way and is about the same size as the within-subject one. But the per-subject differences have a standard deviation of 16.6 points. At $n = 16$ the test had $27\%$ power for the difference observed and would have needed a gap of 12.5 AUC points to reach $80\%$. The cross-subject comparison is uninformative, not contradictory, and the modality claim in this paper is a within-subject claim.

One pattern matters for deployment: ranking transfers across subjects but calibration does not. For one subject the autonomic decoder reached $86.9\%$ AUC at $23.4\%$ accuracy, ordering that subject's epochs well while putting the decision threshold in the wrong place. Any cross-subject deployment would need per-subject recalibration.

\subsubsection{Controls.}
Shuffled labels produced chance performance ($48.9\%$, $p = 0.363$ against 50), so the pipeline does not leak. The motor decoder, using joystick and head motion only, reached $86.2\%$: below the autonomic decoder ($p = 4.9\times10^{-4}$) but far above chance. High accuracy on the task-demand label is therefore not evidence of arousal decoding, which is the point section~\ref{sec:labels} makes from the design side. Shuffled labels are the one destruction control this dataset supports, so the validity evidence in the next section comes instead from comparisons that vary ring size independently of elapsed time.

\subsection{The decoded index is not a proxy for trial phase}\label{sec:validity}

The decoder sees a 2\,s window of physiology and nothing else: no clock, no trial counter, no ring index. Every trace starts at zero by construction, and the smoothing filter damps whatever rise follows, so the opening seconds carry the least information. From the input signals alone, each trial nonetheless recovers the same shape, low at the start and high at the end. That also raises the obvious worry. A signal that ramps within every trial might be reading elapsed time rather than the body. Three tests say it is not.

\subsubsection{Trials do not end where a clock would put them.}
The index ramps within a trial, which is what a clock would do. But a clock makes a further prediction: every trial should end near the top of its own ramp. They do not. Across hard-course trials, $32.4\%$ ended with decoded arousal below 25 and $48.6\%$ below 50, with a standard deviation of 34.6 spanning the full range (figure~\ref{fig:figure_2}A). Mean final arousal by duration quartile is not even monotone (47.7, 39.5, 53.6, 66.6, shortest to longest), and final arousal correlates with trial duration at only $r = +0.225$. Elapsed time alone cannot end half the hard trials at low arousal.

\subsubsection{The index separates the courses at matched trial phase.}
At equal elapsed time the index separated hard from easy trials at $\beta = +6.41$ ($p = 1.1\times10^{-32}$), and the effect strengthened under quadratic control for time ($\beta = +6.71$, $p = 1.6\times10^{-35}$). Restricting to large rings only, which holds the nominal task-demand label fixed, gave $\beta = +6.58$ ($p = 1.4\times10^{-31}$). The coefficient was positive in every time bin. A signal driven purely by trial phase predicts zero.

The comparison signals behave differently (figure~\ref{fig:figure_2}B). At the per-trial scaling used in the original pipeline, the EEG index separated the courses at $\beta = +4.77$ overall but only $\beta = +2.30$ on large rings alone. At per-subject scaling matched to the peripheral index, the same EEG signal gave $\beta = +1.56$ overall and $\beta = -0.73$ on large rings, which is not significant ($p = 0.283$). We report both, because the conclusion for EEG turns on the scaling choice. The motor index was weak and specification-dependent: $\beta = +1.15$ ($p = 0.037$) with linear time control and $\beta = +1.84$ ($p = 0.006$) on large rings, but $\beta = +0.59$ ($p = 0.286$) once time was controlled quadratically, which is where the peripheral effect strengthened. The motor decoder is not a failed control. It discriminates the label at $86.2\%$ AUC and is significant in two of three matched-phase specifications. The peripheral index is four to nine times stronger, and stable across all three.

\subsubsection{Convergent physiological validation.}
\begin{figure}[!htbp]
  \centering
  \includegraphics[width=\linewidth]{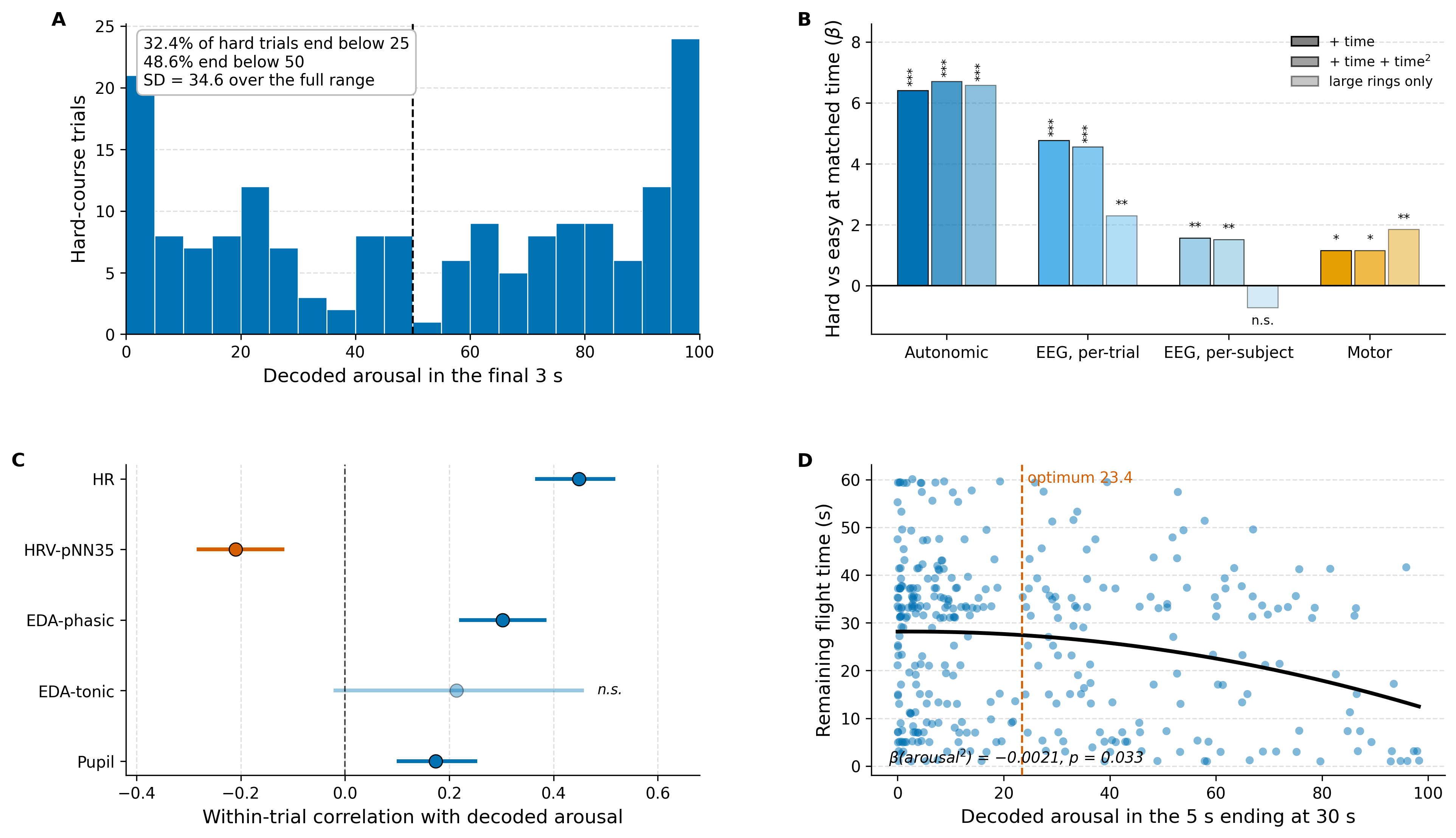}
  \caption{Validity of the decoded index. (A) Distribution of decoded arousal in the final 3\,s of hard-course trials. A signal tracking elapsed time would place every trial near the top of the scale. (B) Course discrimination at matched elapsed time for four indices under three model specifications. Bars are the coefficient for hard versus easy course; stars are $p < 0.05$, $p < 0.01$ and $p < 0.001$. Only the autonomic index is both large and stable across specifications, and the EEG index loses the effect entirely once it is scaled the same way. (C) Within-trial correlation between decoded arousal and each raw physiological channel, averaged across trials, with 95\% confidence intervals across subject means. Tonic EDA is not significant. (D) Landmark analysis: among trials still flying at 30\,s, arousal measured in the 5\,s window ending at 30\,s against the flight time remaining after that point.}
  \label{fig:figure_2}
\end{figure}

Decoded arousal tracked raw autonomic signals in the theoretically expected directions (figure~\ref{fig:figure_2}C). Within-trial correlations averaged across trials were $+0.449$ for HR (95\% CI $[+0.368, +0.515]$, $p = 5.3\times10^{-9}$), $-0.210$ for HRV-pNN35 ($[-0.281, -0.120]$, $p = 1.9\times10^{-4}$), $+0.302$ for phasic EDA ($p = 2.2\times10^{-6}$) and $+0.174$ for pupil diameter ($p = 2.9\times10^{-4}$). Tonic EDA was not significant ($+0.214$, $[-0.019, +0.455]$, $p = 0.091$). The four significant correlations survive Holm correction within this family. The raw traces are min-max scaled within each trial, so any method that reduces a trial to a single mean sees almost no relationship. The choice of method therefore needs stating; supplementary table~S3 compares seven of them. These channels are inputs to the peripheral decoder, so the correlations show physiological interpretability rather than independent validation. This dataset holds no arousal ground truth to validate against.

The motor index also correlates with autonomic physiology, at $+0.324$ with HR, $+0.263$ with phasic EDA and $+0.161$ with pupil. It is an arousal-correlated behavioural signal rather than a non-arousal control, which is unsurprising: effortful stick and head movement covaries with sympathetic arousal in a flight task. The two indices are uncorrelated at trial level ($r = -0.007$), so neither substitutes for the other.

\begin{figure}[!htbp]
  \centering
  \includegraphics[width=\linewidth]{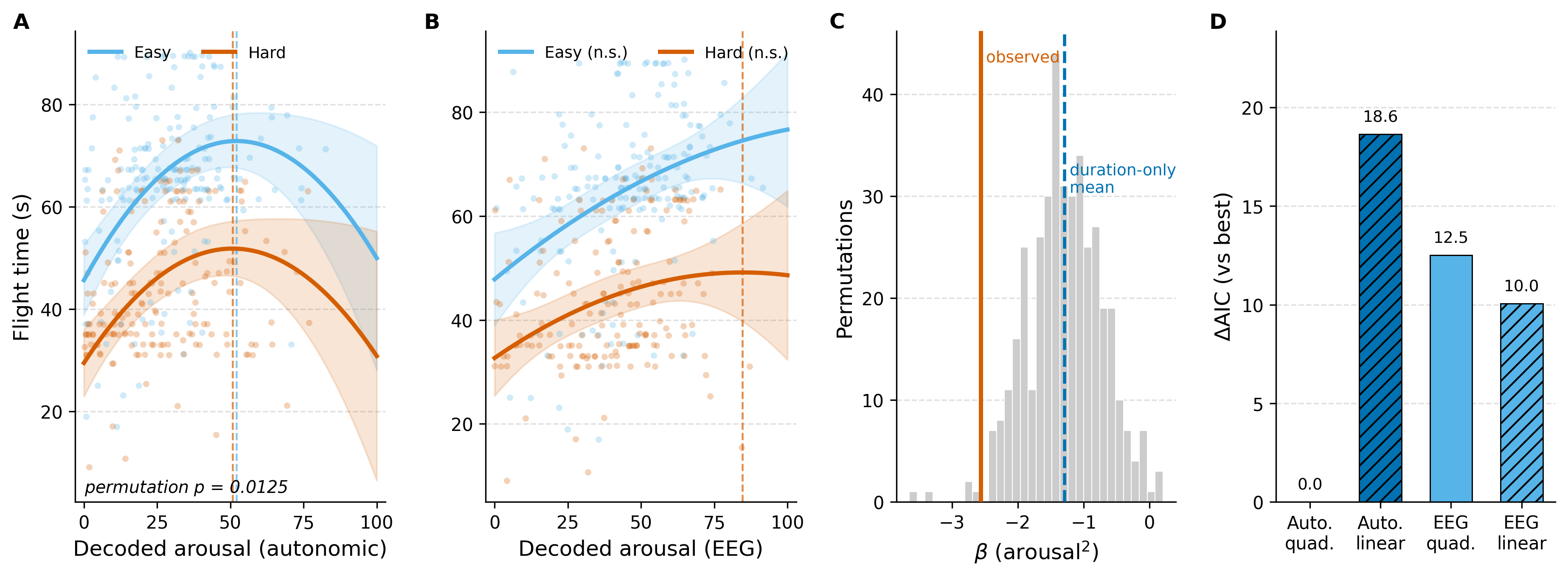}
  \caption{The arousal-performance relationship. (A) Trial-level flight time against mean decoded arousal for the autonomic decoder, with quadratic mixed-model fits by course. Hard course in orange, easy course in blue; shaded bands are 95\% confidence intervals; dashed vertical lines mark the estimated optima (52.2 easy, 50.8 hard). The permutation $p$-value shown is the one that accounts for the coupling between trial-mean arousal and trial length. (B) The same fit for the EEG feedback index rescaled per subject to match the autonomic index. Neither term is significant. (C) Permutation null for the quadratic coefficient, obtained by shuffling arousal traces between trials within duration strata. Duration structure alone accounts for about half of the observed curvature. (D) AIC comparison across model variants relative to the best model.}
  \label{fig:figure_3}
\end{figure}

\subsection{Trial performance follows an inverted-U in decoded arousal}\label{sec:yerkes}

For the autonomic decoder the quadratic term in equation~(\ref{eq:lme}) was negative ($\beta = -2.56$, 95\% CI $[-3.79, -1.33]$), with estimated optima of 52.2 on the easy course and 50.8 on the hard course (figure~\ref{fig:figure_3}A). Because trial-mean arousal is averaged over a window whose length is the outcome, we report the permutation $p$-value rather than the model-based one: the observed coefficient falls below the 400-shuffle null in $1.25\%$ of cases ($p = 0.0125$; figure~\ref{fig:figure_3}C). Duration structure alone produces a mean coefficient of $-1.29$ in that null, so about half the observed curvature comes from the length coupling and only the rest from trial-specific arousal. The same null places the fitted optimum somewhere inside the 0--100 range in $95.5\%$ of shuffles, so an interior peak is on its own no evidence of anything. We therefore treat the two length-safe analyses --- the landmark model and, in particular, the discrete-time hazard model --- as the primary evidence for the inverted-U, with the trial-mean quadratic as corroboration, because only the former separate arousal from trial length by construction.

Model comparison favoured the autonomic quadratic model, with $\Delta\text{AIC} = 18.6$ over the autonomic linear variant and $\Delta\text{AIC} = 12.5$ over the EEG quadratic model at matched scaling (figure~\ref{fig:figure_3}D). The EEG index did not recover the relationship at either scaling ($\beta = -1.30$, $p = 0.097$ at per-trial scaling; $\beta = -0.46$, $p = 0.450$ at per-subject scaling; figure~\ref{fig:figure_3}B), and neither did the motor index at trial level ($\beta = -0.87$, not significant). Of the three indices, only the autonomic one recovers the inverted-U at trial level.

Both length-safe analyses agree. In the landmark analysis at $L = 30$\,s, arousal measured before the outcome window began predicted remaining flight time with a negative quadratic ($\beta = -0.0021$, 95\% CI $[-0.0041, -0.0002]$, $p = 0.033$; figure~\ref{fig:figure_2}D), with an optimum at 23.4; the motor index at the same landmark was null ($p = 0.623$). Only one of six landmarks tested was significant, so this is confirmatory rather than primary. In the hazard model the quadratic on arousal was positive under every control for time, giving a U-shaped hazard, with minimum hazard near arousal 23. Because the 5\,s outcome windows overlap, we report the coefficient with standard errors clustered on trial rather than the model-based ones: $\beta = +1.66$, 95\% CI $[+0.05, +3.28]$, $p = 0.043$. Counting each crash once instead, on non-overlapping 5\,s blocks, weakens it to $p = 0.052$ with the same sign and a minimum near 31 (supplementary section~S9). The instantaneous optimum of 23 to 30 sits well below the trial-mean optimum of about 51. They measure different things, so we keep them separate.

One limb of the curve rests on far less data than the other. With time above and time below the band fitted in a single model, which is why these coefficients differ slightly from table~\ref{tab:deviations}, time above the band predicted worse performance ($\beta = -0.191$, $[-0.286, -0.097]$, $p = 0.0001$) and time below it did not ($\beta = -0.156$, $[-0.482, +0.170]$, $p = 0.349$). The task simply does not produce under-aroused states during demanding phases: only $2.9\%$ of trial time was spent below the band, and $72\%$ of trials never went below it at all. The falling limb is robust across every method used here. The rising limb is not established.

\subsection{Baseline measures from calibration shape the arousal-performance curve}\label{sec:individual}

\begin{figure}[!htbp]
  \centering
  \includegraphics[width=\linewidth]{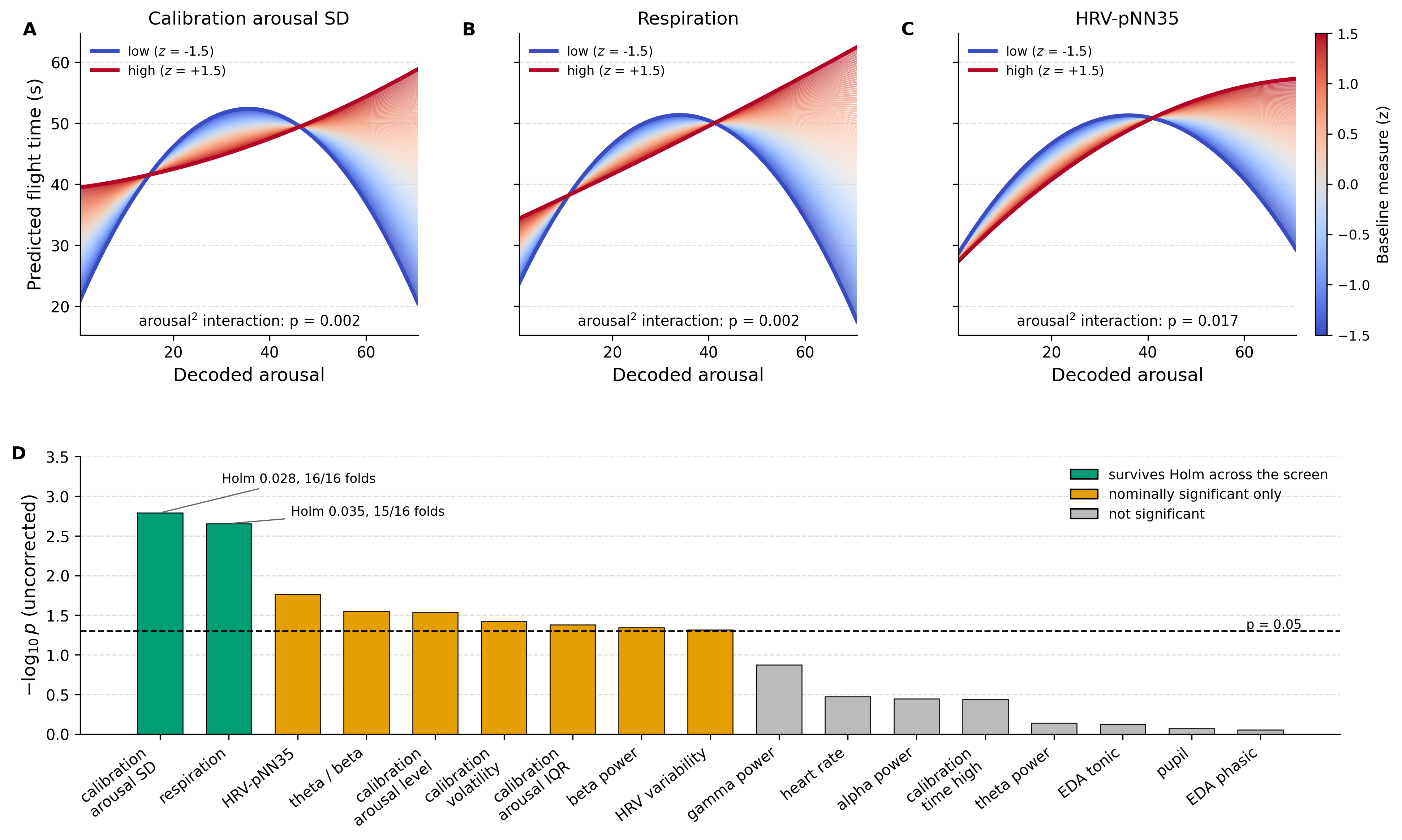}
  \caption{Baseline measures from the calibration block moderate the arousal-performance curve. (A) Predicted flight time against decoded arousal as a function of calibration arousal variability (z-scored), from the mixed-effects model. Curves are drawn over the observed range of trial-mean arousal only. (B) The same for respiration. (C) The same for HRV-pNN35, which was singled out ahead of the screen on theoretical grounds and is reported uncorrected. (D) Screen of all 17 calibration measures against curve-shape moderation, ordered by uncorrected $p$-value. Green bars survive Holm correction across the screen and are annotated with the corrected $p$-value and the number of leave-one-subject-out fits in which they remain significant.}
  \label{fig:figure_4}
\end{figure}

Two of the 17 baseline measures survived Holm correction across the screen (figure~\ref{fig:figure_4}). The variability of the decoded index during calibration moderated the curve most strongly ($\beta = +2.39$, $p = 0.0016$, Holm $p = 0.028$) and stayed significant in all 16 leave-one-subject-out fits (range $p = 0.0005$ to $0.037$). Respiration came second ($\beta = +2.16$, $p = 0.0022$, Holm $p = 0.035$), significant in 15 of 16. Both work the same way: subjects with lower values have a sharply peaked curve that falls steeply at high arousal, and subjects with higher values a flatter one. That is the arousal-sensitive against arousal-tolerant pattern.

HRV-pNN35, singled out ahead of the screen on theoretical grounds (section~\ref{sec:individual_methods}), points the same way ($\beta = +1.099$, SE $0.462$, 95\% CI $[+0.193, +2.005]$, $p = 0.0174$). It is reported uncorrected as a single a priori test. Treated as a member of the screen it would not have survived (Holm $p = 0.26$), and we give both values.

No EEG band survives the screen: gamma is not even nominally significant ($p = 0.134$) and beta only just is ($p = 0.046$). Neither survivor depends on how band power is defined.

The same difference shows up within trials, in a model with a different outcome
and a different unit of analysis. Subjects differ widely in how strongly excess
arousal above the control band raises the chance of crashing within the next
5\,s: adding a random slope on that excess improves the hazard model by
$\chi^2(2) = 78.2$ ($p = 1.1\times10^{-17}$), though only 7 of the 16 subjects spend enough time above the band for that difference to be estimated at all (supplementary section~S9), and per-subject slopes fitted
independently range from $-0.40$ to $+0.013$. The same baseline measures predict that
reactivity, and in the same direction as they predict curve shape. The
interaction between excess arousal and the measure was $\beta = -0.00105$ (95\%
CI $[-0.00161, -0.00049]$, $p = 0.0003$) for respiration, $-0.00068$
($[-0.00114, -0.00021]$, $p = 0.0041$) for calibration arousal variability and
$-0.00045$ ($[-0.00089, -0.00001]$, $p = 0.043$) for HRV. A negative coefficient
means that a higher value of the measure weakens the link between excess arousal
and imminent failure. The two analyses are not independent, but they use
different outcomes and different units, so their agreement is not automatic.
Correlating the 16 independently fitted slopes against each measure directly is
null (all $p > 0.45$), which is what 16 points, one of them extreme, will give
when the pooled model uses every second of data and models the clustering.

Two limits matter for deployment. The measures are reliable within the calibration block (split-half $r$ between $0.65$ and $0.89$), but the calibration \textit{level} of the decoded index cannot carry information forward at all, because the 0--100 scale is defined by that subject's own calibration percentiles. The spread measures survive only because clipping leaves them free to vary. And moderating the curve is a different question from predicting closed-loop behaviour: of 15 tests of the latter, one reached nominal significance and would not survive correction.

Calibration arousal SD was both selected from the 17-measure screen and then used to define the sensitivity split, a select-then-use structure. Two things bound the resulting optimism: it survives Holm correction across the full screen and in all 16 leave-one-subject-out refits, and the split built on it does not itself survive modelling (section~\ref{sec:nostable}), which is why we treat the continuous moderator as primary.

\subsection{What feedback does to the arousal signal}\label{sec:feedback}

\begin{figure}[!htbp]
  \centering
  \includegraphics[width=\linewidth]{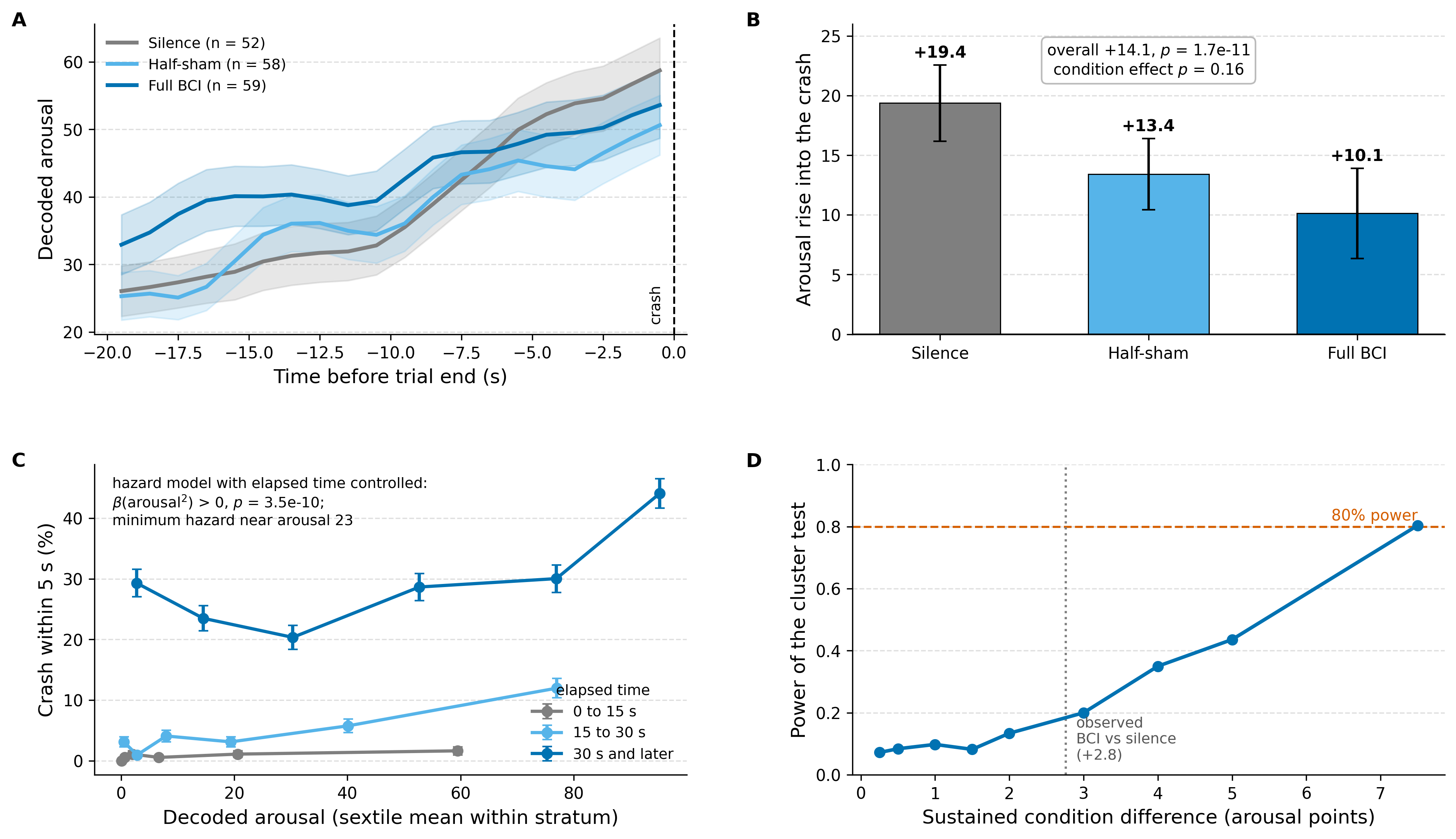}
  \caption{Feedback and the arousal signal. (A) Mean decoded arousal in the 20\,s before the end of each hard-course trial, by feedback condition ($n$ is the number of trials with at least a 20\,s pre-crash window: silence 52, half-sham 58, full BCI 59, of 173 hard-course trials total), with SEM bands. (B) Rise in arousal from the 20 to 10\,s window into the final 10\,s, by condition. The overall rise is highly significant; the difference between conditions is a trend. (C) Probability of crashing within 5\,s against current decoded arousal, computed within elapsed-time strata so that the association is not confounded with trial phase. (D) Power of the cluster-based permutation test against the size of an injected sustained condition difference. The observed BCI against silence difference is marked.}
  \label{fig:figure_5}
\end{figure}

Every one of the 173 hard-course trials ended in a crash rather than at the 90\,s ceiling (longest trial 73\,s), so analysing trial endings selects on nothing. The comparison needs 20\,s of trial before the crash, which 169 of the 173 provide. Mean decoded arousal rose from 33.1 in the window 20 to 10\,s before the crash to 47.2 in the final 10\,s, a rise of 14.1 points (paired $t(168) = 7.23$, $p = 1.7\times10^{-11}$; mixed-model intercept $+13.87$, 95\% CI $[8.31, 19.43]$, $p = 1\times10^{-6}$; figure~\ref{fig:figure_5}A,B). The hazard model gives the same result at finer grain: crash probability within 5\,s rose steeply with current arousal within every elapsed-time stratum (figure~\ref{fig:figure_5}C).

That escalation ordered across feedback conditions as expected, at $+19.4$ points in silence, $+13.4$ under half-sham and $+10.1$ under full BCI, but the difference was not significant ($F = 1.89$, $p = 0.155$; mixed-model likelihood-ratio $p = 0.126$), and we report it as a trend. The hazard model gives a stronger version of the same test. Above-band arousal raised the 5-second crash hazard ($\beta = +0.0498$, $p < 0.001$), and full BCI feedback weakened that coupling (interaction $\beta = -0.0381$, $p = 0.040$). This does not survive correction within the three-test feedback family (Holm $p = 0.080$), so it is suggestive; it is larger and does survive when the band is defined adaptively ($\beta = -0.0535$, $p = 0.002$) or as a sustained excursion of at least 3\,s ($\beta = -0.0530$, $p = 0.007$). Feedback appears to weaken the link between excess arousal and imminent failure rather than to lower arousal overall.

The mean arousal trajectories themselves do not differ. Cluster-based permutation testing found nothing on the easy course, and on the hard course neither the omnibus test across the three conditions ($p = 0.274$) nor the BCI against silence contrast ($p = 0.269$) reached significance. One cluster in the first 7\,s of the BCI against half-sham contrast does not survive correction across the eight trajectory tests (Holm $p = 0.091$), and we do not build on it. A null is only worth the power behind it. The observed BCI minus silence difference was $+2.76$ points with a between-subject SD of $9.69$, and the design reaches $80\%$ power only for sustained offsets of about $7.5$ points (figure~\ref{fig:figure_5}D). We can exclude sustained shifts of $7.5$ arousal points or more, and nothing smaller. Taken with the failure-locked result, feedback looks like it acts near the moments where performance is lost rather than shifting the whole trajectory.

\subsection{Deviation from the optimal trajectory predicts performance}\label{sec:deviation}

\subsubsection{The optimal arousal level is not constant within a trial.}
The optimum read off the joint arousal-performance surface rises monotonically from trial onset to the end of the fitted range (figure~\ref{fig:figure_6}A), as the attentional demand escalates and the rings shrink. A single threshold held constant across trial time would therefore be wrong for much of every trial. The trajectory can be fitted out to about 38\,s, beyond which too few trials survive to populate the surface; figure~\ref{fig:figure_6}C marks the extrapolated region, which enters no statistic.

\subsubsection{Deviation metrics.}
\begin{figure}[!htbp]
  \centering
  \includegraphics[width=\linewidth]{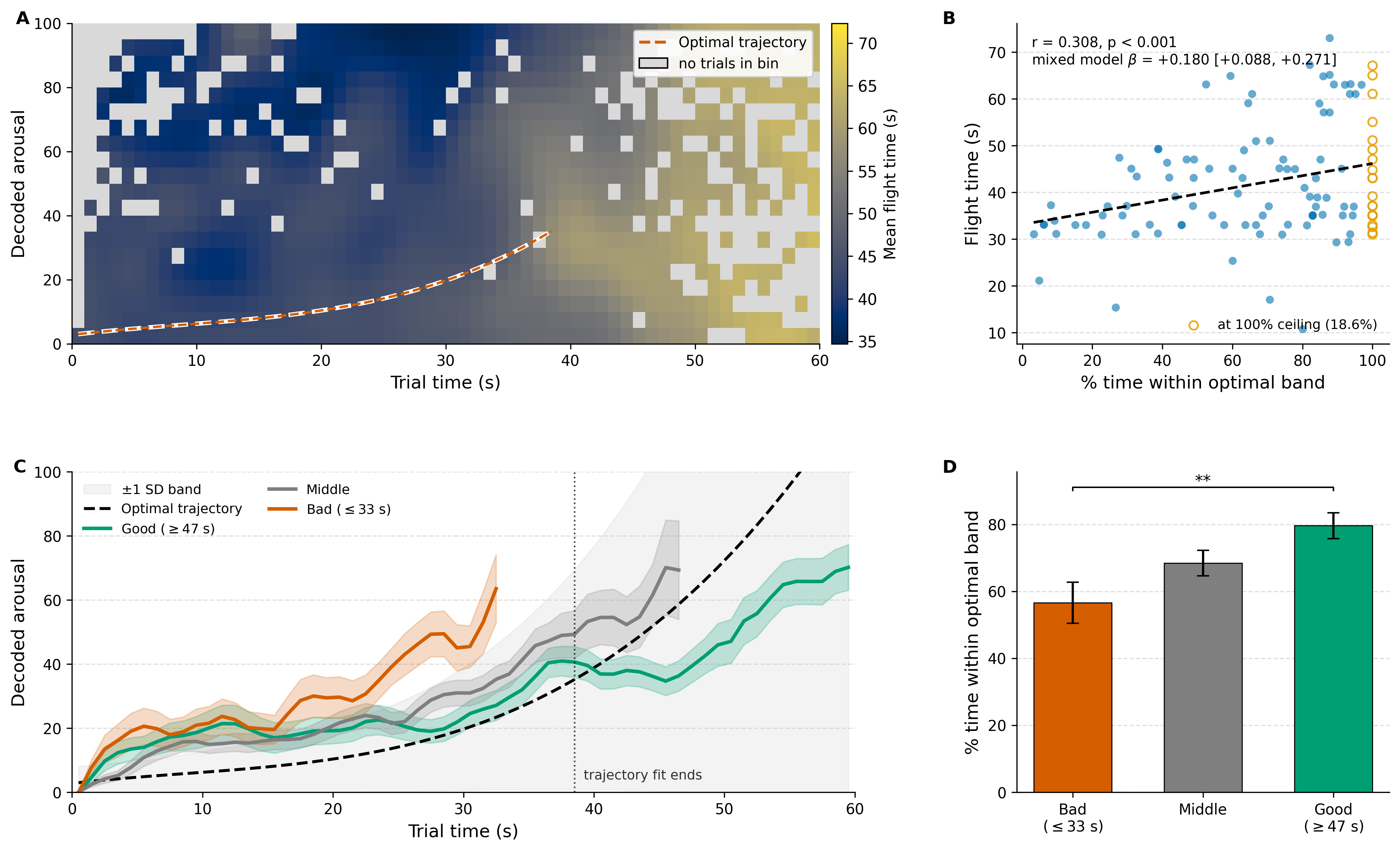}
  \caption{The time-varying optimal arousal trajectory. (A) Mean flight time as a joint function of decoded arousal and trial time, from hard-course control trials. Grey cells contain no trials. The dashed line is the polynomial-fit optimal trajectory over the range in which enough trials survive to fit it. (B) Trial-level flight time against percentage of time within the optimal band. Trials at the $100\%$ ceiling are drawn as open circles. (C) Mean arousal traces with SEM for top-quartile, middle and bottom-quartile trials, over the optimal trajectory and its $\pm 1$ SD band. The dotted line marks where the trajectory fit ends and extrapolation begins. (D) Percentage of time within the band by performance group.}
  \label{fig:figure_6}
\end{figure}

Time spent outside the band predicted shorter flights (table~\ref{tab:deviations}). Under a mixed model with a random intercept per subject, percentage of time above the band gave $\beta = -0.180$ (95\% CI $[-0.271, -0.088]$, $p = 0.0001$) and mean excursion magnitude $\beta = -0.209$ ($[-0.367, -0.051]$, $p = 0.0097$). Trials in the top performance quartile spent $79.6\%$ of their duration within the band against $56.6\%$ for bottom-quartile trials (figure~\ref{fig:figure_6}D; $p < 0.01$, Holm-corrected). The corresponding trajectories separate clearly: poor trials rise above the optimum early, good trials track it from below (figure~\ref{fig:figure_6}C).

Two features of the metrics qualify the table. The distribution is bounded and $18.6\%$ of trials sit exactly at the $100\%$ in-band ceiling, which is why we give Spearman coefficients as well as Pearson. And three of the four metrics defined in section~\ref{sec:individual_methods} are monotone transforms of one another: a bin counts as in-band whenever it is at or below the upper bound, and the excursion rate is the above-band percentage rescaled by trial length. Only mean excursion magnitude is independent, so table~\ref{tab:deviations} has two rows rather than four.

The control-trial set includes the half-sham condition, which delivers half of the veridical feedback signal, and that choice matters. Refitting on true silence trials alone keeps the association in the same direction but reduces it to a trend ($r = +0.251$, $p = 0.072$, $n = 52$). Parameters fitted on silence and applied to half-sham trials give $r = +0.171$ ($p = 0.188$, $n = 61$). Both subsets are small, so these are underpowered rather than contradictory. The dependence on how a control trial is defined should nonetheless be on the record.

A reactive intervention needs time to act, so we measured how long an above-band excursion lasts before arousal comes back down. Bridging dips of $1$\,s or less and discarding excursions shorter than $2$\,s as transients, excursions on poorly performing trials lasted a median of $8$\,s (mean $10.4 \pm 8.5$\,s, $n = 35$), or $6$\,s at the retuned band. That is the latency budget a reactive controller has to fit inside, measured here rather than assumed.

\begin{table}[htbp]
  \caption{Deviation metrics against flight time for hard-course control trials (113 trials, 16 subjects). Coefficients are from linear mixed models with a random intercept per subject; Pearson and Spearman coefficients are descriptive. Percentage of time within the band and excursion rate are monotone transforms of the first row and are not reported separately.}
  \label{tab:deviations}
  \centering
  \begin{tabular}{lccc}
    \hline
    Metric & Pearson / Spearman & $\beta$ [95\% CI] & $p$ \\
    \hline
    \% time above band       & $-0.308$ / $-0.248$ & $-0.180$ $[-0.271, -0.088]$ & $0.0001$ \\
    Mean excursion magnitude & $-0.254$ / $-0.227$ & $-0.209$ $[-0.367, -0.051]$ & $0.0097$ \\
    \hline
  \end{tabular}
\end{table}

\subsection{Individualization: what is real and what is not}\label{sec:personalisation}

\subsubsection{Held-out control bands.}\label{sec:heldout_bands}
Every band scheme below is scored on the same 113 held-out trials, with the trajectory, the group split and every multiplier refitted on the 15 training subjects in each fold (figure~\ref{fig:figure_7}A). These correlations are therefore not the ones in table~\ref{tab:deviations}, which fits the trajectory on all subjects at once: the default band gives $r = +0.298$ here against $+0.308$ there. The benchmark matters as much as the schemes do. A personalized band has to beat a universal band retuned to its best single multiplier, which uses no per-subject information at all, so we report that alongside the default. Supplementary tables~S6 and~S7 give every scheme and every contrast.

We tested four personalized schemes. Giving each half of the sample its own band width, split at the median of the calibration measure that survives the screen, reached $r = +0.320$ and $d = 0.83$, with stable multipliers across folds ($1.75$ to $2.25$ for the sensitive half, $1.5$ to $3.0$ for the tolerant half). Deriving each subject's width from their own fitted curve instead of by grid search reached $r = +0.307$ and $d = 0.83$, with multipliers from $0.62$ to $1.59$ (median $1.01$). Moving the band \textit{centre} to each subject's own fitted optimum beat both, at $r = +0.328$ and $d = 0.95$, the best of any scheme on both measures; the shifts were within a few arousal units for 14 of 16 subjects. Combining the shifted centre with the derived width was worse than the centre alone ($r = +0.317$).

None of them separates from the benchmark. The retuned universal band reaches $r = +0.328$ and $d = 0.91$ by itself (pooled across held-out subjects; scored within subject, as in section~\ref{sec:bandwidth}, the same band gives $r = +0.349$), against the default band's $+0.298$ ($\Delta r = +0.031$, 95\% CI $[-0.002, +0.062]$, $p = 0.064$), and every personalized scheme falls inside its noise: group widths give $\Delta r = -0.007$ ($p = 0.752$) against it, the shifted centre $\Delta r = -0.000$ ($[-0.049, +0.069]$, $p = 0.907$). Against the default instead, the shifted centre gives $\Delta r = +0.031$ ($[-0.021, +0.111]$, $p = 0.407$). With 113 trials from 16 subjects the study cannot resolve differences this small. Carrying the fitted parameters unchanged onto the held-out full-BCI condition settles nothing either: $r = +0.169$ ($p = 0.198$) for the universal band and $r = +0.187$ ($p = 0.152$) for the personalized one, at $n = 60$ and power around $0.35$.

Band centre is the one per-subject parameter that had not been varied before, and the only one that improved held-out discrimination on both measures. That makes it a direction worth testing, not a result. The individual differences behind it are established (section~\ref{sec:individual}); what a properly powered study still has to show is that acting on them beats a retuned global policy.

\subsubsection{There is no stable individual target to personalize on.}\label{sec:nostable}
The group split describes a difference that does not survive modelling. Pooled within each half, the tolerant subjects show $r = +0.415$ and the sensitive subjects $r = +0.228$, which is the reverse of the ordering the sensitivity account predicts. Nothing else supports the contrast in either direction. The group-by-time-in-band interaction is not significant ($\beta = +0.057$, $p = 0.538$; likelihood-ratio $p = 0.510$; AIC worsens from 869.0 to 871.7). The continuous version is null ($p = 0.193$). And fitting each subject separately gives mean slopes of $+0.304$ for the eight sensitive subjects against $+0.273$ for the seven tolerant subjects with enough trials ($t = +0.18$, $p = 0.861$). The pooled contrast is a description, not a moderation.

The problem runs deeper than an unstable median split (figure~\ref{fig:figure_7}B). Estimated independently in random halves of a subject's own trials, that subject's optimal arousal looks reliable: $r = +0.70$ for the top-quartile estimator and $+0.89$ for the performance-weighted one. But both estimators correlate with the subject's plain mean arousal at $+0.94$ and $+0.99$, and mean arousal, which uses no performance information at all, is itself reliable at $r = +0.92$. Residualize the optimum on that mean and the reliability collapses to $+0.02$ and $+0.14$, with intervals spanning zero. There is no reliable person-specific optimum beyond where that person's index sits on average, and per-subject rescaling already normalizes level. Band width is no better. Split-half agreement across a subject's own trials is $\rho = +0.06$ ($p = 0.84$) by rank and $r = +0.27$ ($p = 0.33$) by product moment, the same multiplier is selected in $7\%$ of subjects, and the mean disagreement is $0.92$ on a grid spanning $0.5$ to $3.0$. Subjects are not identical, and adding a random slope on arousal does improve the model ($\chi^2(2) = 6.52$, $p = 0.038$), but that would not survive the correction policy and a random slope on arousal$^2$ does not converge.

\subsubsection{Band width is the wrong control parameter.}\label{sec:bandwidth}
The natural response to a negative result is that a better predictor would help. It would not, because the trouble is the parameter, not what sets it. Band width is a constant within a subject, while the association between deviation and performance lives inside subjects: decomposed, it is $r = +0.353$ within subject ($p < 0.0001$) and $r = +0.190$ between subjects ($p = 0.48$, $n = 16$). A constant cannot reorder a subject's own trials. A sixfold change in width leaves the trial ordering correlated at $r = +0.886$ with the universal band. What it does change is each subject's mean, which injects between-subject variance into any pooled statistic. Widening the band also pins a rising share of trials at the $100\%$ ceiling, from $21.4\%$ at $m = 1.0$ to $43.4\%$ at $m = 3.0$, discarding the discrimination the metric depends on. Amplifying or inverting the mapping from calibration measure to multiplier changes nothing (supplementary table~S8).

There is real headroom, but it is not reachable. An oracle picking each subject's best multiplier in sample does gain when scored within subject, on the metric it optimizes (mean within-subject $r = +0.409$ against $+0.330$ for the universal band, $p = 0.0068$), and that gain does not cross-validate (leave-one-trial-out $r = +0.363$, $p = 0.46$). This comparison is within-subject while the held-out figures above are pooled, so the universal band takes a different value in each; only the figures in section~\ref{sec:heldout_bands} compare across schemes. Most of the honest improvement comes from retuning the \textit{universal} multiplier to $m = 1.75$ (within-subject $r = +0.349$, the same scoring as the in-sample oracle it is compared against here; the pooled held-out value for this band is $r = +0.328$, section~\ref{sec:heldout_bands}), with no personalization at all.

\subsubsection{A control parameter that does add information.}
\begin{figure}[!htbp]
  \centering
  \includegraphics[width=\linewidth]{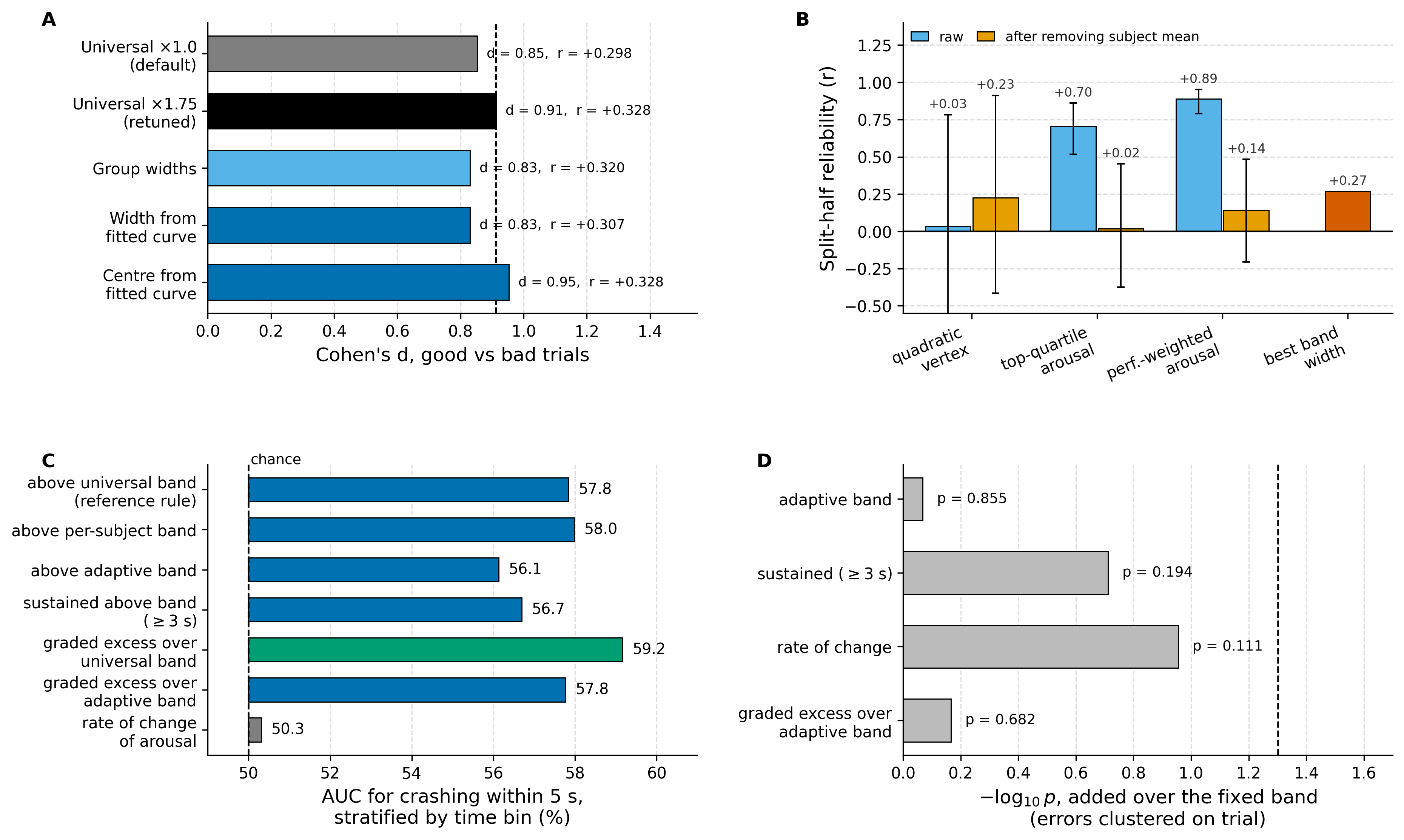}
  \caption{Individualization, evaluated out of sample. (A) Five control bands scored on the same 113 held-out trials, with the trajectory, the group split and every multiplier refitted on the training subjects in each fold. Bars are Cohen's d between good and bad trials; the correlation between time in band and flight time is printed beside each. The dashed line marks the retuned universal band, which uses no per-subject information and is the benchmark a personalized scheme has to beat. No scheme separates from it. (B) Split-half reliability of three estimators of a subject's optimal arousal, before and after removing that subject's own mean arousal, with 95\% ranges across 200 random splits. Nothing survives removal of the mean. The band-width bar carries no such range, being a single coefficient across the 16 subjects; the rank coefficient for it is $\rho = +0.06$. (C) Discrimination of candidate control parameters for crashing within 5\,s, computed within elapsed-time strata. The graded excess over the universal band, in green, discriminates best. (D) Whether each parameter adds information over the fixed universal band, with standard errors clustered on trial. None does. The linear probability fit used at submission put the graded excess above an adaptively referenced band at $p = 6.6\times10^{-4}$ (supplementary section~S9).}
  \label{fig:figure_7}
\end{figure}

If the useful variation is within a trial, the control parameter has to vary within the trial too. In the hazard framework (figure~\ref{fig:figure_7}C,D), one candidate did: the graded distance above a band whose width tracks the subject's own arousal variability in a trailing 10\,s window added information over the graded excess above the fixed band ($\beta = +0.0020$, likelihood-ratio $\chi^2 = 11.6$, $p = 6.6\times10^{-4}$). A binary adaptive threshold did not ($p = 0.77$), nor did a sustained-excursion rule ($p = 0.13$). The rate of change of arousal was nominally significant ($p = 0.026$) but fails correction within this family (Holm $p = 0.078$) and discriminates at chance alone (AUC 50.3), so we drop it. What matters, then, is how far above the band a subject is, measured against a reference that tracks their current variability, rather than whether they are above it at all. Taken alone the adaptive reference is not even the strongest signal: graded excess above the fixed band discriminates slightly better (AUC $59.2$ against $57.8$; figure~\ref{fig:figure_7}C). But it is the only candidate carrying information the fixed rule does not, which is what a controller needs from an extra input. That advantage does not hold once the overlapping windows are accounted for: the two graded signals correlate at $r = 0.913$, and entered together the fixed excess keeps the effect while the adaptive one does not (supplementary section~S9). What survives is the graded form rather than the reference.

One measurement caution goes with this result. Taken marginally, being above the band predicts imminent failure at chance (AUC $49.9$), even though the mixed-model coefficient is highly significant ($p = 7\times10^{-10}$). Both numbers are correct. Late in a trial, elapsed time pushes the outcome and the band signal in opposite directions: the fitted trajectory rises, so almost nothing exceeds the band after about 45\,s, while the crash rate stays high, and the two effects cancel. Stratified by time bin, the same indicator gives AUC $57.8$ and raises the 5\,s crash rate by 7.0 percentage points, positive in 7 of 9 bins. The stratified figure is the one that answers the controller's question.

% ═══════════════════════════════════════════════════════════════════════════════
\section{Discussion}

\subsection{The case for peripheral signals in arousal-based BCIs}

Within subject, the peripheral decoder beats EEG by roughly 8 AUC points and the FBCSP benchmark by 13, which is a large margin by BCI standards. Across subjects the same comparison settles nothing: the difference has the same sign and a similar size, but a confidence interval spanning zero at $27\%$ power. This paper therefore supports one claim about modality, that peripheral signals are the stronger candidate within subject in this task, and not the general one.

The reason is straightforward. The LC-NE system \cite{astonjones2005} drives cortical excitability and peripheral sympathetic-parasympathetic balance together, and the cardiac channels are among the most direct readouts of that state, so their dominance in the attribution analysis is what one would expect. Arousal is not something that happens in cortex and is echoed by the body. It is a state the body expresses, and the scalp is one more place to look for it. Peripheral sensors also sidestep a systematic risk in EEG-based arousal decoding: the gamma-band EMG confound identified in Faller et al \cite{faller2019}, where scalp muscle artefact mimics high-frequency cortical activity and inflates apparent arousal sensitivity. Adding EEG to the autonomic decoder gains almost nothing ($+1.2$ AUC points, $p = 0.016$, with near-zero attribution in figure~\ref{fig:figure_1}D), so once the autonomic axis is represented, scalp EEG carries little extra information about graded arousal here. That is a statement about this task. EEG may retain value in other cognitive and affective BCI contexts, and nothing here speaks to those.

\subsection{What the validity evidence does and does not establish}\label{sec:validity_discussion}

The decoder is trained on a task-demand label that is almost perfectly determined by position within a trial, so accuracy cannot settle whether it decodes arousal or phase, and neither can label manipulation in this dataset. Three results speak to it. The index does not behave like a clock: half of hard-course trials end at low arousal, and final arousal is barely related to trial duration. It separates the two courses at matched elapsed time, including when the nominal label is held constant, where a matched-scaling EEG index does not. And it tracks HR, HRV, phasic EDA and pupil diameter in the directions the LC-NE account predicts. Together these show that the index carries trial-specific physiological information beyond elapsed time. They do not show that it measures arousal in the sense a pharmacological or thermal manipulation would, and no analysis of this dataset could.

Recovery of the inverted-U should be read the same way. Only one of the three indices recovers the relationship at trial level, which says something about the comparison. It is also a single paradigm with ring size as the demand manipulation, which limits what it says about arousal decoders generally. The arousal-performance relationship need not be a stable inverted-U across tasks, individuals, stressors and time scales, and learning, fatigue, strategy change and failure events can all bend the curve. So we do not propose Yerkes-Dodson compliance as a general validity criterion. It is proof-of-concept evidence in one paradigm, alongside matched-phase discrimination and convergent physiological evidence, and it would have to be tested across paradigms and modalities before it could serve as a standard.

The motor decoder is easy to misread in either direction, so it deserves a plain statement. It reaches $86.2\%$ AUC on the task-demand label and correlates with HR at $+0.32$, phasic EDA at $+0.26$ and pupil at $+0.16$. It is not a non-arousal control: effortful stick and head movement covaries with sympathetic arousal in a flight task, and its U-shaped crash hazard is what a noisy arousal proxy should give, not a counterexample. What separates the two indices is degree and stability. The peripheral index tracks autonomic physiology about 1.4 times more strongly, discriminates course at matched phase four to nine times more strongly, and does so stably across specifications. It recovers the trial-level inverted-U where the motor index does not, and the two are uncorrelated at trial level.

\subsection{Reframing what neurofeedback does}

Arousal escalates sharply into failure, by 14.1 points in the final 10\,s, and full BCI feedback weakens the coupling between above-band arousal and imminent failure. That weakening is only suggestive under the fixed band definition and clearer under the adaptive and sustained ones. Meanwhile sustained shifts in the overall trajectory of $7.5$ arousal points or more can be excluded, and smaller ones cannot. The natural assumption, that auditory neurofeedback works by lowering average arousal across a trial, gets no support from these data, though the design could not have detected a modest version of it.

This changes what to measure and what to build. Judging a closed-loop arousal intervention by its effect on mean or peak arousal will tend to underestimate it; performance-conditional analyses, such as arousal at the moment of failure, are more sensitive. And an intervention precise enough to act at critical moments is a better deployment target than one that biases arousal throughout a trial. We cannot characterise those local effects beyond the failure-locked window here. Higher-resolution analyses around specific failure events are the natural follow-up, and this dataset is underpowered for them.

\subsection{Individualization as a design problem}

The individualization results are mostly negative, and the negatives are the more useful half. What is real is a measure available before the loop closes: the variability of the decoded index during the calibration block. It moderates the shape of the arousal-performance curve, survives correction across a 17-measure screen, and stays significant when any single subject is dropped. Respiration behaves the same way, and HRV points the same way as a single a priori test. No EEG band power survives the screen.

What is not real is a per-subject control target. Remove a subject's own mean and no reliable person-specific optimum is left, and a subject's best band width cannot be estimated reliably from a single session. That explains the group results without blaming an unstable median split, and it carries a concrete design implication: estimating a person-specific optimum precisely enough to beat that person's own mean takes repeated sessions. One session and roughly eleven hard-course trials per subject will not do it.

The individual differences themselves are not in doubt. They show up at trial
level as a change in the shape of the arousal-performance curve and within
trials as a change in how strongly excess arousal predicts imminent failure, in
two models with different outcomes and different units of analysis. People
differ in what excess arousal costs them, and ten minutes of calibration
predicts it.

Band width turns out to be the wrong knob, for a structural reason no better predictor can fix. The association lives within subjects, a constant cannot reorder a subject's own trials, and an in-sample oracle's advantage does not cross-validate; most of the reachable gain comes from retuning the universal multiplier instead. The matching positive is specific. A graded distance above the band, evaluated every second, predicts imminent failure where a binary threshold barely does. The negative and the positive are one finding seen from two sides: the informative variation is within the trial, so the control parameter has to be as well.

Band centre is the one per-subject parameter we varied that improved held-out
discrimination on both measures. We do not claim it works. Its confidence
interval spans zero, and the deeper problem is the benchmark: a universal band
retuned to one multiplier for everybody reaches the same correlation using no
per-subject information at all, and nothing we tested separates from it. A
personalized scheme scored against the default band rather than a retuned one
will always look better than it is, which is why we report both. What the
finding justifies is a prediction a larger study could test directly, with a
preregistered comparison, a retuned global band as the control, and enough
subjects to resolve a difference in correlation of about 0.03. Given a
between-subject SD of the size seen here, that is a matter of subject count and
trials per subject rather than of finding a better marker.

\subsection{Toward a closed-loop implementation}\label{sec:framework}

None of what follows was tested here. The peripheral decoder was never in the feedback loop and no intervention was ever delivered on the basis of it, so the components below are a specification drawn from offline analysis.

A system built on these results would work as follows. A 10-minute calibration session trains the decoder and records the variability of its own output, which adds no experimental burden and is already part of most BCI paradigms. Peripheral signals then drive the decoder in real time at 62.5\,ms resolution, giving a per-subject normalized index. That index is compared against a time-varying optimal trajectory with a universal band width, retuned to $m = 1.75$ rather than personalized, and the trigger is the graded distance above a band whose width tracks the subject's own arousal variability in a trailing window. Calibration variability enters as a stratification variable, for analysis and for setting expectations about a given operator, not as a band-width parameter: scaling width by it does not improve control.

Could such a system act in time? Sustained above-band excursions on poorly performing trials lasted a median of 8\,s at the default band and 6\,s at the retuned one. That is a workable latency budget for a reactive intervention, and it is measured here rather than imported from the stimulation literature.

\subsection{Future work}

The moment-targeted picture in section~\ref{sec:feedback} suggests a specific comparison. Auditory neurofeedback asks the operator to interpret and act on a signal while flying a high-workload task. Transcutaneous vagus nerve stimulation bypasses that attentional bottleneck. It recruits an afferent pathway whose central projections have been mapped with fMRI \cite{frangos2015, yakunina2017}, it reduces sympathetic nerve activity in healthy humans \cite{clancy2014}, and it changes heart rate during stimulation trains as short as 60\,s \cite{badran2018}. Cardiac vagal effects on heart rate begin under 1\,s after efferent traffic changes and peak within a few seconds \cite{berntson1997}, which the dwell-time budget above can accommodate. The end-to-end latency of a stimulation loop in this setting has not been measured, so that budget is a design constraint rather than a demonstrated fit. Comparing the two modalities within an otherwise identical closed-loop architecture would test whether the bottleneck in auditory feedback lies in the precision of self-regulation or in the latency of the physiological response.

A prospective study would differ from Faller et al \cite{faller2019} in three ways. It would use a decoder with higher within-subject accuracy, whose output has been checked against trial phase and against raw autonomic signals. Its trigger would be graded deviation from the band rather than a threshold crossing. And its stimulation modality would match the moment-targeted hypothesis. A preregistered crossover design (active tVNS, sham tVNS, silence) with $N \approx 24$ subjects gives $80\%$ power on the Faller et al effect size alone. We do not assume additional variance reduction from personalization, since the present results do not support it.

The individual-difference results point to two further studies. The first is a
band-centre comparison. Differences in reactivity to excess arousal are not
resolved here, and setting the band centre from each subject's own fitted
curve is the most promising way of acting on them that we found, but the
improvement was not significant at $N = 16$. Testing it properly needs enough
subjects and enough trials per subject to resolve a difference in the
deviation-performance correlation of roughly 0.03, and a retuned universal band
as the comparison rather than the default one. The second follows from the
reliability analysis: pinning down a per-subject arousal optimum precisely
enough to personalize against needs repeated sessions, because one session
cannot estimate it any better than that subject's own mean arousal.

\subsection{Limitations}

The sample is 16 subjects in a single session, and the individualization analyses divide it further. Group-specific multipliers come from a grid search on a noisy objective with few degrees of freedom, and the small gain in the held-out evaluation has a bootstrap interval that does not survive correction. Splitting subjects at the median also discards graded information and puts an unstable boundary through a small sample, which is one reason we treat the continuous moderation results as primary. Confirming the stability of the calibration measures and any group-specific parameters needs independent replication in 30 to 40 subjects.

Other constraints come from the paradigm rather than the sample. The task-demand label is close to a deterministic function of trial phase, which caps what accuracy on it can show, leaves a phase-stratified evaluation undefined and makes any label-shift control uninformative. The rising limb of the inverted-U rests on very little data, because the task does not produce under-aroused states during demanding phases, though the falling limb is robust across every method used here. The control-trial set includes half-sham trials, and restricting to true silence reduces the deviation-performance association to a trend. Tonic EDA does not reach significance in the convergent physiological analysis, and $18.6\%$ of trials sit at the ceiling of the band metric.

Cross-subject generalization is limited in a way that matters for deployment. Leave-one-subject-out accuracy is 16.8 points below within-subject accuracy, with three subjects near chance, and the ordering of subjects transfers while the decision threshold does not, so per-subject recalibration would be required.

Some signal-processing choices were fixed a priori rather than tuned, among them the pNN35 cardiac feature, which departs from the more common pNN50. Sweeping the filter constants and the decoder warm-up leaves the results unchanged, and removing the smoothing entirely changes nothing that matters, but we did not sweep the cardiac feature itself, and a systematic comparison against pNN50 is warranted.

Finally, the decoder was trained on one paradigm with ring size as the demand manipulation. Whether it generalizes to other sources of arousal, such as sustained fatigue, emotional stressors or pharmacological intervention, is unknown. And because the Faller et al experiment ran EEG-based neurofeedback rather than the peripheral index evaluated here, nothing in these data attributes their performance benefit to feedback quality. Establishing that would take a prospective study.

% ═══════════════════════════════════════════════════════════════════════════════
\section{Conclusion}

Arousal is a state of the whole body, and reading it from the body works. Within subject, peripheral signals decode task-demand-related arousal more accurately than EEG in a demanding sensorimotor task, and the resulting index carries trial-specific information that elapsed time cannot explain. Trials end across the full range of the scale. The index separates the two courses at matched trial phase, where a matched-scaling EEG index does not, and tracks heart rate, heart rate variability, electrodermal activity and pupil diameter in the expected directions. Of the three indices compared, only the peripheral one recovers an inverted-U with trial performance. Across subjects the modality comparison is too weak to confirm or deny the same ordering, so the claim is a within-subject one.

Arousal escalates sharply in the seconds before failure, and veridical feedback weakens the link between excess arousal and imminent failure, while sustained shifts in the overall trajectory larger than about 7.5 arousal points can be excluded. One measure available before the loop closes shapes the arousal-performance curve: the variability of the decoded index during calibration, selected from a 17-measure screen and still to be confirmed prospectively. Scaling control-band width by it, or by any other per-subject measure, does not improve control, and we show why: the association lives within trials, so a per-subject constant cannot exploit it. A graded distance above the band can, though which reference sets that band is not settled here. Together these results specify a candidate architecture for individualized closed-loop arousal regulation and name the control parameter it should act on, as a target for prospective validation rather than a system that has been validated.

% ─── BACK MATTER ──────────────────────────────────────────────────────────────

\section*{Funding}
This work was supported by funding from the Army Research Laboratory’s Human-Guide Intelligent Systems (HGIS) Program (W911NF-23-2-0067), the Army Research Laboratory’s STRONG Program (W911NF-19-2-0139, W911NF-19-2-0135, W911NF-21-2-0125),  the Air Force Office of Scientific Research (FA9550-22-1-0337), and a Vannevar Bush Faculty Fellowship from the US Department of Defense (N00014-20-1-2027). The funders played no role in study design, data collection, analysis and interpretation of data, or the writing of this manuscript.

\section*{Data availability}
Original data from Faller et al \cite{faller2019} are publicly available via IEEE DataPort (\url{https://doi.org/10.21227/rn3e-bp31}).

\section*{Code availability}
All analysis code is openly available at \url{https://github.com/anirudhmn/ArousalDecoding} under the MIT licence. The repository contains the decoder implementation, the training pipeline, the scripts that produce every figure and table in this paper and its supplementary material, and one standalone script per validation, control and individual-difference analysis. A driver (\texttt{analyses/run\_all.sh}) runs them in dependency order and writes a log for each. Model fitting is seeded per subject, repetition and fold. Four scripts retrain decoders and take about five hours in total. Their per-subject accuracies are committed as tables, so the decoding results can be checked without retraining anything, and a second driver (\texttt{analyses/run\_fast.sh}) reruns the remaining twenty-six analyses in about ten minutes. Every log and every summary table in the repository was produced by the run that the reported numbers come from. The raw dataset is obtained separately from IEEE DataPort at the link given above.

\section*{Ethical statement}
The analyses in this manuscript use a previously collected, publicly available dataset (Faller et al., 2019). The original data collection was approved by the Institutional Review Board (IRB) of Columbia University, and written informed consent was obtained from all participants prior to the experiment.

\section*{Conflict of interest statement}
The authors declare that they have no known competing financial interests or personal relationships that could have appeared to influence the work reported in this paper.

% ═══════════════════════════════════════════════════════════════════════════════
% SUPPLEMENTARY MATERIAL
\clearpage
\appendix
\setcounter{section}{0}
\setcounter{table}{0}
\renewcommand{\thetable}{S\arabic{table}}
\renewcommand{\thesection}{S\arabic{section}}
\begin{center}
{\Large\bfseries Supplementary material}
\end{center}
Every table here is written by a script in the analysis repository
(\url{https://github.com/anirudhmn/ArousalDecoding}), named at the end of each
section. Running \texttt{analyses/run\_all.sh} regenerates all of them from the
public dataset.

\section{Registry of every inferential $p$-value}

Table~\ref{tab:registry} lists the 75 $p$-values that support a claim in the
main text, the family each belongs to, and the Holm-corrected value within that
family. A family is a set of tests addressing one question with interchangeable
outcomes. Descriptive quantities carry no $p$-value and do not appear. Nor do
tests that qualify a claim rather than support one: the alternative
specifications of the matched-phase model, the refits on narrower control-trial
subsets, the in-sample oracle, the within-versus-between decomposition, the six
candidate landmarks and the fifteen calibration-to-closed-loop transfer tests.
Each of those is marked where it appears.

The table makes three things easier to see than the main text can. Of the 75
tests, 45 are significant uncorrected and 33 survive their family's correction,
so 12 nominally significant results are reported as not surviving: the feedback
interaction on the hazard, the one trajectory cluster, and every calibration
measure below the top two. HRV-pNN35 appears twice, once as the single a priori test
motivated by Thayer et al, uncorrected, and once inside the 17-measure
screen, where it does not survive; both are shown. And the four rows of the Table~1 family are
monotone transforms of one another, so correcting them as four independent tests
is conservative. The main text reports two of them and says so.

\begingroup
\footnotesize
\setlength{\tabcolsep}{4pt}
\begin{longtable}{p{3.4cm}p{5.6cm}rrcp{1.1cm}}
\caption{Every inferential $p$-value, its family, and the Holm-corrected value
within that family. ``Survives'' is the corrected value against
$\alpha = 0.05$.}\label{tab:registry}\\
\toprule
Family & Test & $p$ & Holm $p$ & Survives & Note \\
\midrule
\endfirsthead
\multicolumn{6}{l}{\textit{Table \ref{tab:registry} continued}}\\
\toprule
Family & Test & $p$ & Holm $p$ & Survives & Note \\
\midrule
\endhead
\bottomrule
\endfoot
decoder comparisons & peripheral vs shuffled & 1.95e-14 & 1.75e-13 & yes & \\
decoder comparisons & LOSO peripheral vs chance & 3.70e-07 & 2.96e-06 & yes & \\
decoder comparisons & combined vs EEG (within-subject) & 4.20e-07 & 2.96e-06 & yes & \\
decoder comparisons & LOSO EEG vs chance & 1.10e-06 & 6.60e-06 & yes & \\
decoder comparisons & peripheral vs EEG (within-subject) & 8.70e-06 & 4.35e-05 & yes & \\
decoder comparisons & within-subject vs LOSO & 1.30e-05 & 5.20e-05 & yes & \\
decoder comparisons & peripheral vs motor & 4.90e-04 & 0.0015 & yes & \\
decoder comparisons & combined vs peripheral (within-subject) & 0.0160 & 0.0320 & yes & \\
decoder comparisons & LOSO peripheral vs EEG & 0.1726 & 0.1726 & no & \\
\midrule
label controls & real vs reversed (79.5\% changed) & 3.88e-04 & 0.0019 & yes & \\
label controls & real vs phase-perm (0.8\% changed) & 0.0480 & 0.1920 & no & \\
label controls & real vs block-perm (0.0\% changed) & 0.0780 & 0.2340 & no & \\
label controls & real vs cross-trial (0.9\% changed) & 0.0952 & 0.2340 & no & \\
label controls & real vs shift2 (6.7\% changed) & 0.3350 & 0.3350 & no & \\
\midrule
validity & matched-phase, $+$ time$^2$ & 1.59e-35 & 4.77e-35 & yes & \\
validity & matched-phase, all crossings $+$ time & 1.07e-32 & 2.14e-32 & yes & primary \\
validity & matched-phase, large rings only & 1.40e-31 & 1.40e-31 & yes & \\
\midrule
Yerkes-Dodson & hazard quadratic, 5\,s-bin fixed effects & 3.50e-10 & 1.05e-09 & yes & primary, see~S9 \\
Yerkes-Dodson & trial-mean quadratic, duration permutation & 0.0125 & 0.0250 & yes & corroboration \\
Yerkes-Dodson & landmark analysis at $L = 30$\,s & 0.0328 & 0.0328 & yes & \\
\midrule
Table 1 & \% time above band & 1.00e-04 & 4.00e-04 & yes & \\
Table 1 & \% time in band & 1.00e-04 & 4.00e-04 & yes & \\
Table 1 & excursion rate & 1.00e-04 & 4.00e-04 & yes & \\
Table 1 & mean excursion magnitude & 0.0097 & 0.0097 & yes & \\
\midrule
a priori (family of one) & arousal$^2$ $\times$ HRV & 0.0174 & 0.0174 & yes & primary \\
\midrule
calibration screen (17) & calibration arousal SD & 0.0016 & 0.0272 & yes & \\
calibration screen (17) & respiration & 0.0022 & 0.0352 & yes & \\
calibration screen (17) & HRV & 0.0174 & 0.2610 & no & \\
calibration screen (17) & theta/beta & 0.0280 & 0.3920 & no & \\
calibration screen (17) & calibration level & 0.0293 & 0.3920 & no & \\
calibration screen (17) & calibration volatility & 0.0381 & 0.4572 & no & \\
calibration screen (17) & calibration IQR & 0.0420 & 0.4620 & no & \\
calibration screen (17) & beta & 0.0456 & 0.4620 & no & \\
calibration screen (17) & HRV SD & 0.0485 & 0.4620 & no & \\
calibration screen (17) & gamma & 0.1341 & 1.0000 & no & \\
calibration screen (17) & heart rate & 0.3361 & 1.0000 & no & \\
calibration screen (17) & alpha & 0.3577 & 1.0000 & no & \\
calibration screen (17) & calibration frac-high & 0.3611 & 1.0000 & no & \\
calibration screen (17) & theta & 0.7223 & 1.0000 & no & \\
calibration screen (17) & EDA tonic & 0.7586 & 1.0000 & no & \\
calibration screen (17) & pupil & 0.8356 & 1.0000 & no & \\
calibration screen (17) & EDA phasic & 0.8840 & 1.0000 & no & \\
\midrule
group moderation & continuous marker $\times$ \%in-band & 0.1925 & 0.5775 & no & primary \\
group moderation & median-split group $\times$ \%in-band & 0.5380 & 1.0000 & no & \\
group moderation & per-subject slopes, sensitive vs tolerant & 0.8610 & 1.0000 & no & \\
\midrule
held-out control bands & retuned universal vs default & 0.0640 & 0.4480 & no & \\
held-out control bands & width from fitted curve vs default & 0.2920 & 1.0000 & no & \\
held-out control bands & width from fitted curve vs retuned & 0.3440 & 1.0000 & no & \\
held-out control bands & group widths vs default & 0.3540 & 1.0000 & no & \\
held-out control bands & centre from fitted curve vs default & 0.4070 & 1.0000 & no & \\
held-out control bands & group widths vs retuned & 0.7520 & 1.0000 & no & \\
held-out control bands & centre from fitted curve vs retuned & 0.9070 & 1.0000 & no & \\
\midrule
individual reactivity & random slope, excess above adaptive band & 1.06e-17 & 5.30e-17 & yes & \\
individual reactivity & random slope, excess above fixed band & 4.02e-10 & 1.61e-09 & yes & \\
individual reactivity & excess $\times$ respiration & 3.00e-04 & 9.00e-04 & yes & \\
individual reactivity & excess $\times$ calibration arousal SD & 0.0041 & 0.0082 & yes & \\
individual reactivity & excess $\times$ HRV & 0.0429 & 0.0429 & yes & \\
\midrule
within-trial control & graded excess over adaptive band & 6.56e-04 & 0.0026 & yes & primary \\
within-trial control & rate of change of arousal & 0.0261 & 0.0783 & no & \\
within-trial control & sustained excursion ($\geq 3$\,s) & 0.1280 & 0.2560 & no & \\
within-trial control & binary adaptive band & 0.7700 & 0.7700 & no & \\
\midrule
reliability & random slope on arousal (LRT) & 0.0384 & 0.0768 & no & \\
reliability & band-width split-half agreement & 0.8430 & 0.8430 & no & \\
\midrule
feedback effects & failure-locked rise, paired $t$ & 1.70e-11 & 5.10e-11 & yes & primary \\
feedback effects & above-band $\times$ full BCI on 5\,s hazard & 0.0400 & 0.0800 & no & \\
feedback effects & failure-locked rise by condition (ANOVA) & 0.1550 & 0.1550 & no & \\
\midrule
trajectory clusters & hard: BCI vs half-sham & 0.0228 & 0.0912 & no & \\
trajectory clusters & hard: BCI vs half-sham (2nd cluster) & 0.2209 & 0.6627 & no & \\
trajectory clusters & hard: BCI vs silence & 0.2690 & 0.6627 & no & \\
trajectory clusters & hard: omnibus & 0.2741 & 0.6627 & no & \\
\midrule
convergent physiology & HR & 5.34e-09 & 2.67e-08 & yes & \\
convergent physiology & EDA-phasic & 2.24e-06 & 8.96e-06 & yes & \\
convergent physiology & HRV-pNN35 & 1.93e-04 & 5.79e-04 & yes & \\
convergent physiology & Pupil & 2.88e-04 & 5.79e-04 & yes & \\
convergent physiology & EDA-tonic & 0.0910 & 0.0910 & no & \\
\end{longtable}
\endgroup

\noindent Script: \texttt{analyses/multiplicity.py}.

% ─────────────────────────────────────────────────────────────────────────────
\section{The 17-measure calibration screen}

Table~\ref{tab:screen} gives every measure entered into the screen described in
section 3.5 of the main text, with the coefficient on the
$\text{Measure}_z \times \text{Ar}^2$ interaction. A positive coefficient means
that subjects with higher values of the measure have a flatter
arousal-performance curve. The last column counts the leave-one-subject-out
refits in which the measure remained significant at $p < 0.05$, which is
reported only for the two measures that survive the screen.

The four EEG band powers are mean square band-passed amplitude. Beta is
nominally significant and gamma is not, and neither survives correction.
Recomputing band power as variance, a Hilbert envelope, a Welch spectral density
or log power changes nothing here. The two survivors are unaffected either way,
since respiration and the decoder-derived measures have one definition each.

\begin{table}[htbp]
\caption{Baseline measures screened for moderation of the arousal-performance
curve. Holm correction is across all 17 members.}\label{tab:screen}
\centering
\footnotesize
\begin{tabular}{llrrrc}
\toprule
Measure & Source & $\beta$ & $p$ & Holm $p$ & LOSO fits significant \\
\midrule
calibration arousal SD & decoder & $+2.390$ & 0.0016 & 0.028 & 16/16 \\
respiration & physiological & $+2.162$ & 0.0022 & 0.035 & 15/16 \\
HRV-pNN35 & physiological & $+1.099$ & 0.0174 & 0.261 & -- \\
theta/beta ratio & physiological & $-2.546$ & 0.0280 & 0.392 & -- \\
calibration arousal level & decoder & $-2.236$ & 0.0293 & 0.392 & -- \\
calibration volatility & decoder & $-1.592$ & 0.0381 & 0.457 & -- \\
calibration arousal IQR & decoder & $+1.880$ & 0.0420 & 0.462 & -- \\
beta power & physiological & $+1.050$ & 0.0456 & 0.462 & -- \\
HRV variability & physiological & $-1.373$ & 0.0485 & 0.462 & -- \\
gamma power & physiological & $+0.674$ & 0.1341 & 1.000 & -- \\
heart rate & physiological & $-0.752$ & 0.3361 & 1.000 & -- \\
alpha power & physiological & $+1.264$ & 0.3577 & 1.000 & -- \\
calibration time high & decoder & $-0.813$ & 0.3611 & 1.000 & -- \\
theta power & physiological & $+0.535$ & 0.7223 & 1.000 & -- \\
EDA tonic & physiological & $+0.203$ & 0.7586 & 1.000 & -- \\
pupil diameter & physiological & $+0.162$ & 0.8356 & 1.000 & -- \\
EDA phasic & physiological & $-0.096$ & 0.8840 & 1.000 & -- \\
\bottomrule
\end{tabular}
\end{table}

\noindent Script: \texttt{analyses/calibration\_profile.py}.

% ─────────────────────────────────────────────────────────────────────────────
\section{Seven ways of correlating the index with raw physiology}

The main text reports the within-trial correlation between decoded arousal and
each raw physiological channel, averaged across trials. That choice is not
neutral, and the seven alternatives in table~\ref{tab:corrmethods} differ enough
that it is worth stating plainly.

The trial-mean and subject-mean rows collapse to nothing because of the
pipeline, not the physiology. The raw traces are min-max scaled within each
trial, so every trial starts at 0 and ends at 100 by construction, and any
method that reduces a trial to one number has already thrown away the
between-trial variation it needs. The within-trial methods see shape, which is
what survives that scaling. The pooled-sample row treats 4.8 million samples as
independent and is shown only to confirm that the estimate is not an artefact of
averaging.

\begin{table}[htbp]
\caption{The same relationship computed seven ways. Values are Pearson
coefficients. The row used in the main text is marked.}\label{tab:corrmethods}
\centering
\footnotesize
\begin{tabular}{lrrrrrc}
\toprule
Method & HR & HRV-pNN35 & EDA-phasic & EDA-tonic & Pupil & Unit of analysis \\
\midrule
Within-trial, averaged (reported) & $+0.449$ & $-0.210$ & $+0.302$ & $+0.214$ & $+0.174$ & 352 trials \\
Within-trial, Fisher $z$ & $+0.519$ & $-0.250$ & $+0.348$ & $+0.334$ & $+0.192$ & 352 trials \\
Pooled samples & $+0.388$ & $-0.211$ & $+0.288$ & $+0.195$ & $+0.149$ & 4.8M samples \\
Trial means & $+0.003$ & $-0.087$ & $+0.052$ & $-0.083$ & $+0.101$ & 352 trials \\
Subject means & $-0.000$ & $+0.072$ & $-0.096$ & $-0.365$ & $+0.143$ & 16 subjects \\
Within-subject across trials & $+0.034$ & $-0.194$ & $+0.206$ & $+0.095$ & $+0.106$ & 16 subjects \\
Mixed model, partial $r$ & $+0.002$ & $-0.182$ & $+0.170$ & $+0.031$ & $+0.055$ & 352 trials \\
\bottomrule
\end{tabular}
\end{table}

Significance for the reported row is assessed with a one-sample test across
subject means, with a rank check alongside. HR gives $+0.449$, 95\% CI
$[+0.368, +0.515]$, $p = 5.3\times10^{-9}$, Spearman $+0.453$. HRV-pNN35 gives
$-0.210$, $[-0.281, -0.120]$, $p = 1.9\times10^{-4}$, Spearman $-0.225$.
EDA-phasic gives $+0.302$, $[+0.222, +0.383]$, $p = 2.2\times10^{-6}$, Spearman
$+0.334$. EDA-tonic gives $+0.214$, $[-0.019, +0.455]$, $p = 0.091$, Spearman
$+0.131$. Pupil gives $+0.174$, $[+0.103, +0.250]$, $p = 2.9\times10^{-4}$,
Spearman $+0.160$.

\noindent Script: \texttt{analyses/external\_validation.py}.

% ─────────────────────────────────────────────────────────────────────────────
\section{Why a label-shift control cannot be built in this dataset}

Temporally shifted labels are the standard control for separating a
physiological decoder from a phase decoder. Table~\ref{tab:labels} shows why the
control is uninformative here rather than merely inconvenient. Ring size
decreases on a fixed schedule within a trial, so the task-demand label is close
to a deterministic function of position within the trial. Ring position alone
predicts the label at 99.8\% AUC within subject, the mapping is exactly
deterministic for 13 of the 16 subjects, and only 18 of 854 ring-position strata
contain both classes.

Every label manipulation that respects trial structure therefore changes almost
no labels. Shifting by two ring positions changes 6.7\% of them, permuting
within phase 0.8\%, and permuting within block none at all. A decoder retrained
on labels that are 99\% identical must score about the same, so the resulting
AUC says nothing about what the decoder uses. Only reversing the labels changes
enough of them to matter, and that control does separate from the real labels.

\begin{table}[htbp]
\caption{Label controls. ``Changed'' is the percentage of labels that differ
from the real ones. AUC is the mean across 16 subjects.}\label{tab:labels}
\centering
\footnotesize
\begin{tabular}{lrrr}
\toprule
Label condition & Labels changed & Mean AUC & $p$ vs real \\
\midrule
Real labels & 0.0\% & 93.37 & -- \\
Shift by two ring positions & 6.7\% & 92.77 & 0.335 \\
Cross-trial permutation & 0.9\% & 92.56 & 0.095 \\
Permutation within phase & 0.8\% & 92.66 & 0.048 \\
Permutation within block & 0.0\% & 93.23 & 0.078 \\
Reversed & 79.5\% & 90.35 & 3.9e-04 \\
\bottomrule
\end{tabular}
\end{table}

The controls that do work are reported in the main text: shuffled labels give
chance performance (48.9\%), and a decoder using only joystick and head motion
reaches 86.2\%, which is the result that limits what accuracy on this label can
demonstrate.

\noindent Script: \texttt{analyses/label\_controls.py}.

% ─────────────────────────────────────────────────────────────────────────────
\section{Sensitivity to the smoothing filter}

The decoded index is smoothed with an asymmetric first-order IIR filter,
$\tau_\text{rise} = 0.75$\,s and $\tau_\text{fall} = 1.5$\,s, chosen a priori to
match sympathetic activation against parasympathetic recovery rather than tuned
on the data. That filter is invertible: each stored trace holds one value per
16-sample update, and the direction of each step reveals which time constant was
applied, so the pre-filter sequence is recoverable exactly and can be re-smoothed
with any other constants without retraining a decoder. A round trip through the
inversion and the primary filter returns the original trace to within
$10^{-13}$ arousal units.

Table~\ref{tab:filter} applies four alternatives, including no smoothing at all.
Every row recomputes trial-mean arousal from the resmoothed trace, so the
comparison across rows is like for like, and the first row differs slightly from
the headline quadratic in the main text, which uses the stored trial means.
Nothing changes sign, no conclusion turns on the filter, and removing the
smoothing entirely leaves the deviation-performance correlation marginally
stronger than the primary value.

\begin{table}[htbp]
\caption{Downstream results under five smoothing choices.}\label{tab:filter}
\centering
\footnotesize
\begin{tabular}{lrrrrrr}
\toprule
Filter & Mean arousal & $r$(\% above, flight) & Cohen's $d$ & $\beta$ quad & $p$ & Optimum \\
\midrule
Asymmetric 0.75 / 1.5 (primary) & 28.71 & $-0.308$ & 0.85 & $-2.333$ & $<0.0001$ & 51.6 \\
Symmetric 1.0 & 26.93 & $-0.310$ & 0.84 & $-2.438$ & $<0.0001$ & 50.1 \\
Symmetric 1.5 & 26.37 & $-0.273$ & 0.74 & $-2.503$ & $<0.0001$ & 49.9 \\
Symmetric 0.75 & 27.22 & $-0.300$ & 0.83 & $-2.405$ & $<0.0001$ & 50.2 \\
No smoothing & 28.07 & $-0.314$ & 0.86 & $-2.303$ & $<0.0001$ & 50.5 \\
\bottomrule
\end{tabular}
\end{table}

\noindent Script: \texttt{analyses/filter\_sensitivity.py}.

% ─────────────────────────────────────────────────────────────────────────────
\section{Every control-band scheme on the same held-out trials}

Table~\ref{tab:bands} scores five control bands on the same 113 held-out
hard-course control trials. In every fold the optimal trajectory, the group
split and every multiplier are refitted on the training subjects, and the
held-out subject is scored with parameters it contributed nothing to.

Table~\ref{tab:banddeltas} bootstraps each difference in correlation at the
subject level, against two baselines: the default universal band at one standard
deviation, and a universal band retuned to its best single multiplier, which
uses no per-subject information at all and is the comparison a personalized
scheme has to win. No scheme separates from the retuned band, and the largest
contrast against the default belongs to the retuned band itself.

\begin{table}[htbp]
\caption{Held-out performance of five control bands. Correlations are pooled across held-out subjects. The retuned universal band scored within subject instead, as in the oracle comparison of the main text, gives $r = +0.349$.}\label{tab:bands}
\centering
\footnotesize
\begin{tabular}{lrrrr}
\toprule
Scheme & $n$ & $r$ & $p$ & Cohen's $d$ \\
\midrule
Universal $\times 1.0$ (default) & 113 & $+0.298$ & 0.0013 & 0.85 \\
Universal $\times 1.75$ (retuned, no personalisation) & 113 & $+0.328$ & 0.0004 & 0.91 \\
Group widths, split at the calibration marker & 113 & $+0.320$ & 0.0006 & 0.83 \\
Centre from fitted curve & 113 & $+0.328$ & 0.0004 & 0.95 \\
Width from fitted curve & 113 & $+0.307$ & 0.0009 & 0.83 \\
\bottomrule
\end{tabular}
\end{table}

\begin{table}[htbp]
\caption{Subject-level bootstrap of each difference in correlation, 10{,}000
resamples.}\label{tab:banddeltas}
\centering
\footnotesize
\begin{tabular}{lrrr}
\toprule
Contrast & $\Delta r$ & 95\% CI & $p$ \\
\midrule
Retuned universal vs default & $+0.031$ & $[-0.002, +0.062]$ & 0.064 \\
Group widths vs default & $+0.023$ & $[-0.028, +0.073]$ & 0.354 \\
Centre from fitted curve vs default & $+0.031$ & $[-0.021, +0.111]$ & 0.407 \\
Width from fitted curve vs default & $+0.010$ & $[-0.009, +0.029]$ & 0.292 \\
\midrule
Group widths vs retuned & $-0.007$ & $[-0.050, +0.027]$ & 0.752 \\
Centre from fitted curve vs retuned & $-0.000$ & $[-0.049, +0.069]$ & 0.907 \\
Width from fitted curve vs retuned & $-0.020$ & $[-0.062, +0.024]$ & 0.344 \\
\bottomrule
\end{tabular}
\end{table}

Selecting the multipliers and scoring them on the same trials inflates the
result. Chosen in sample by grid search on the very trials they are then scored
on, the two group multipliers give $r = +0.337$ against $+0.308$ for a universal
band, a gain of $+0.030$; held out, the same scheme gains $+0.023$, and both
intervals include zero. The in-sample figure is reported only to show the size
of the selection effect and supports no claim.

\noindent Scripts, all in \texttt{analyses/}: \texttt{band\_scheme\_comparison.py},\\
\texttt{loso\_control\_bands.py} and \texttt{individual\_reactivity.py}.

% ─────────────────────────────────────────────────────────────────────────────
\section{Why band width is the wrong control parameter}

Section 4.7 of the main text argues that no calibration measure can rescue
band-width personalization, because the parameter has almost nothing to act on.
Table~\ref{tab:widthlimits} gives the two sweeps behind that claim.

The first sweep varies the mapping from calibration measure to multiplier. The
proportional mapping is the one a controller would use; amplifying it two- and
fourfold widens the range of multipliers it produces, and the inverted mappings
reverse its sign. Nothing moves. Amplifying fourfold leaves the correlation at
$+0.268$ against $+0.270$ for a universal band, and inverting makes it worse.
The knob turns and the outcome does not, which rules out a timid or a backwards
mapping.

The second sweep varies the universal multiplier itself and shows why. Across a
sixfold change in width, from $m = 0.5$ to $m = 3.0$, the trial ordering stays
correlated at $r = +0.886$ or better with the ordering at $m = 1.0$. Widening
the band also pins more and more trials at the $100\%$ in-band ceiling, from
$16.2\%$ at $m = 0.5$ to $43.4\%$ at $m = 3.0$, throwing away the
discrimination the metric depends on.

\begin{table}[htbp]
\caption{Two sweeps over band width. Upper: the mapping from calibration
measure to multiplier, amplified and inverted. Lower: the universal multiplier
itself, with the correlation of the resulting trial ordering against the
ordering at $m = 1.0$.}\label{tab:widthlimits}
\centering
\footnotesize
\begin{tabular}{lrrr}
\toprule
Mapping & Multiplier range & $r$(\% in band, flight) & Cohen's $d$ \\
\midrule
Universal $\times 1.0$ & $1.00$ & $+0.270$ & 0.84 \\
Proportional to the calibration measure & $0.50$--$1.44$ & $+0.269$ & 0.84 \\
Amplified $\times 2$ & $0.50$--$1.89$ & $+0.264$ & 0.81 \\
Amplified $\times 4$ & $0.50$--$2.78$ & $+0.268$ & 0.82 \\
Inverted, amplified $\times 2$ & $0.50$--$2.15$ & $+0.250$ & 0.82 \\
Inverted, amplified $\times 4$ & $0.50$--$3.00$ & $+0.236$ & 0.78 \\
\midrule
Universal multiplier $m$ & Mean \% in band & $r$ vs ordering at $m = 1.0$ & \% at ceiling \\
\midrule
$0.50$ & $64.3$ & $+0.982$ & $16.2$ \\
$1.00$ & $70.4$ & $+1.000$ & $21.4$ \\
$1.75$ & $77.9$ & $+0.968$ & $33.5$ \\
$2.50$ & $82.6$ & $+0.917$ & $40.5$ \\
$3.00$ & $84.8$ & $+0.886$ & $43.4$ \\
\bottomrule
\end{tabular}
\end{table}

The decomposition in the main text explains both sweeps. The
deviation-performance association is $r = +0.353$ within subject
($p < 0.0001$) and $r = +0.190$ between subjects ($p = 0.48$, $n = 16$). A
per-subject multiplier is constant within a subject, so it can act only on the
between component, and that is not where the association lives.

\noindent Script: \texttt{analyses/band\_width\_limits.py}.

% ─────────────────────────────────────────────────────────────────────────────
\section{Cluster-based permutation parameters}

The permutation unit is the subject, with condition labels permuted within
subject, because trials are nested within subjects. The observation is one mean
trajectory per subject per condition, so each subject contributes equally
regardless of trial count. Subjects missing any condition are dropped, which
leaves 16 subjects on the hard course and 15 on the easy course.

The pointwise statistic is a repeated-measures $F$ across the three conditions
and a paired $t$ for pairwise contrasts. The cluster-forming threshold is the
$p < 0.05$ critical value of that statistic: $F = 3.316$ and $t = 2.131$ on the
hard course, $F = 3.340$ and $t = 2.145$ on the easy course. The cluster
statistic is the sum of the pointwise statistic over contiguous supra-threshold
bins, and the null distribution is the maximum cluster statistic per
permutation, which controls the family-wise error rate across time. There are
10{,}000 permutations over a 30\,s window binned at 1\,s, of which the first two
bins are dropped. The first second of every trace is exactly zero by
construction, because sliding-window inference starts at sample 256, and the
filter is still ramping through the second. Neither bin has any between-subject
variance.

On the hard course, the omnibus test gives one cluster at 4 to 6\,s with mass
7.24 and $p = 0.274$. Full BCI against silence gives one cluster at 17 to 19\,s
with mass 4.61 and $p = 0.269$. Full BCI against half-sham gives a cluster at 0
to 7\,s with mass 17.79 and $p = 0.023$, which does not survive Holm correction
across the eight trajectory tests ($p = 0.091$), and a second cluster at 8 to
10\,s with $p = 0.221$. Half-sham against silence gives no supra-threshold
cluster. The easy course gives no supra-threshold cluster in any contrast.

The power curve for the hard-course BCI against silence contrast is figure 5D of
the main text. The observed difference is $+2.76$ arousal points with a
between-subject standard deviation of subject means of 9.69. Power reaches 0.80
only for sustained offsets of about 7.5 arousal points, so that is the size the
design can exclude.

\noindent Script: \texttt{analyses/cluster\_permutation.py}.

\section{Dependence between overlapping outcome windows}\label{sec:glmm}

The hazard analyses take the 1\,s bin as the unit and ask whether the trial
crashes within the next 5\,s. Two features of that need handling. The outcome is
binary and was fitted with a linear mixed model, which is a linear probability
approximation. And the 5\,s windows overlap, so a single crash marks five
consecutive rows and the model counts it five times. Bins are also nested within
trials, which the published bin tables carry no identifier for.

The overlap is a property of the question rather than a mistake in it. A
controller evaluates every second, so consecutive decisions overlap by
construction, and the every-second framing is the one a deployed system would
face. What the overlap breaks is the standard error. Querying the model more
often does not create more crashes, and there are 173 in these data however
often it is queried.

Two refits follow from that. The first keeps the published outcome and bins
exactly as they are and fits them as a binomial GEE with sandwich standard
errors clustered on trial, so the question is unchanged and only the dependent
rows stop counting as independent. The second counts each crash once: the trial
is cut into consecutive non-overlapping 5\,s blocks, the outcome is whether it
ends inside the block, and the predictor is the state at the start of the block,
which keeps the same 0 to 5\,s lead. Two further specifications say how far the
answer travels. A person-period model on 2\,s intervals scores the event only
where the trial ends, which asks about the instant of the crash rather than the
next five seconds. A lagged model asks whether arousal now predicts a crash five
seconds later. In the person-period model the cluster is the subject, because
each trial contributes exactly one event by construction and a within-trial
working correlation is then degenerate, and the baseline hazard uses collapsed
time strata, because 9 of the 173 events fall in the first 30\,s across two
thirds of the time at risk and the first 5\,s stratum separates under any
binomial link.

\begin{table}[htbp]
\centering
\footnotesize
\caption{The quadratic on arousal across specifications. The joint test is of
both arousal terms together.}\label{tab:glmm}
\begin{tabular}{p{6.2cm}rp{3.0cm}rr}
\toprule
Specification & $\beta_2$ & 95\% CI & $p$ & joint $p$ \\
\midrule
published linear probability fit & --- & --- & 3.5e-10 & --- \\
\midrule
\textbf{crash-within-5\,s, GEE on trial clusters} & $+1.66$ & $[+0.05, +3.28]$ & 0.043 & --- \\
non-overlapping 5\,s blocks & $+1.96$ & $[-0.02, +3.93]$ & 0.052 & 0.0016 \\
5\,s blocks, block mean & $+1.83$ & $[+0.01, +3.64]$ & 0.049 & 0.0095 \\
5\,s blocks, logit & $+2.37$ & $[-0.02, +4.76]$ & 0.052 & 0.0025 \\
person-period, instant of the crash & $+0.96$ & $[-0.68, +2.59]$ & 0.251 & 0.0031 \\
5\,s lead & $+1.14$ & $[-0.71, +2.99]$ & 0.228 & 0.0001 \\
\bottomrule
\end{tabular}
\end{table}

The curvature holds with the dependence handled and the question unchanged
($p = 0.043$). Counting each crash once weakens it to $p = 0.052$ with the same
sign and a minimum near arousal 31 rather than 23. It fades once the question
narrows to the instant of the crash, which is a different and stricter question
than the one the paper asks. The arousal effect itself is significant in every
specification that reports a joint test.

One consequence for table~S1. That table records each test as specified, so the
hazard row carries $3.5\times10^{-10}$ and is the smallest $p$ in the
Yerkes-Dodson family, which is why all three tests in that family survive Holm
correction. Substituting $0.043$ reorders the family, and the landmark analysis
then corrects to $0.066$ rather than $0.033$. Neither the landmark analysis nor
the hazard quadratic would survive correction on that substitution, though both
remain significant uncorrected.

\begin{table}[htbp]
\centering
\footnotesize
\caption{The other two conclusions Reviewer~1 asked about, as published and
refitted.}\label{tab:glmm2}
\begin{tabular}{p{5.6cm}p{4.2cm}p{4.6cm}}
\toprule
Conclusion & Published & Refitted \\
\midrule
above band raises the hazard & $\beta = +0.0498$, $p < 0.001$ & HR $1.72$, $[1.06, 2.81]$, $p = 0.029$ \\
above band $\times$ full BCI & $\beta = -0.0381$, $p = 0.040$ & $\chi^2(2) = 1.12$, $p = 0.57$ \\
subjects differ in reactivity & $\chi^2(2) = 78.2$, $p = 1.1\times10^{-17}$ & not estimable, see below \\
respiration moderates reactivity & $\beta = -0.00105$, $p = 0.0003$ & $p = 0.22$ \\
adaptive band adds information & $\beta = +0.0020$, $p = 6.6\times10^{-4}$ & $\beta = -0.0026$, $p = 0.80$ \\
\bottomrule
\end{tabular}
\end{table}

Two of these need a word. The adaptive band reverses rather than weakens. The
two graded signals correlate at $r = 0.913$, and entered together the fixed
excess keeps the effect ($p = 0.039$) while the adaptive one does not. No
candidate adds information over the fixed rule once the dependence is handled.
Graded excess over the fixed band still predicts failure on its own
($p = 0.014$), so what survives is the graded form rather than the reference
that sets the band.

The reactivity difference between subjects cannot be estimated at this sample
size. A subject's own slope on excess arousal needs blocks in which that subject
was above the band, and only 7 of the 16 have 20 or more. Below that the
per-subject logistic quasi-separates and returns converged but meaningless
coefficients, one of them $-44.9$ against a pooled effect of $+0.02$. Restricted
to the seven with enough data the slopes run from $-0.029$ to $+0.075$, and a
permutation test on their spread gives between $p = 0.001$ and $p = 0.16$
depending on whether spread is measured as the standard deviation, the
interquartile range or the median absolute deviation. We report the difference
as untested rather than absent.

\noindent Script: \texttt{analyses/hazard\_glmm.py}.

\end{document}